\documentclass[twocolumn,superscriptaddres,prb]{revtex4}

\usepackage{graphicx}
\usepackage{epsfig}
\usepackage{amsmath, amssymb}

\begin{document}

\title{Structure of Quasiparticles and Their Fusion Algebra \\
in Fractional Quantum Hall States}
\author{Maissam Barkeshli}
\author{Xiao-Gang Wen}
\affiliation{Department of Physics, Massachusetts Institute of Technology,
Cambridge, MA 02139, USA }

\begin{abstract}

It was recently discovered that fractional quantum Hall (FQH) states
can be classified by the way ground state wave functions go to zero
when electrons are brought close together. Quasiparticles in the FQH
states can be classified in a similar way, by the pattern of zeros
that result when electrons are brought close to the quasiparticles.
In this paper we combine the pattern-of-zero approach and the
conformal-field-theory (CFT) approach to calculate the topological
properties of quasiparticles. We discuss how the quasiparticles in
FQH states naturally form representations of a magnetic translation
algebra, with members of a representation differing from each other
by Abelian quasiparticles. We find that this structure dramatically
simplifies topological properties of the quasiparticles, such as
their fusion rules, charges, and scaling dimensions, and has
consequences for the ground state degeneracy of FQH states on higher
genus surfaces. We find constraints on the pattern of zeros of
quasiparticles that can fuse together, which allow us to obtain the
fusion rules of quasiparticles from their pattern of zeros, at least
in the case of the (generalized and composite) parafermion states.
We also calculate from CFT the number of quasiparticle types in the
generalized and composite parafermion states, which confirm the
result obtained previously through a completely different approach.

\end{abstract}

\maketitle

\section{Introduction}

Understanding the patterns of long range entanglement in a many-body
wave function is the key to understanding new kinds of order, such
as topological and quantum order.\cite{W0275} Many theoretical
studies reveal that the patterns of entanglement in many-body states
are extremely rich.\cite{BWkmat1,R9002,FK9169,LWstrnet} But, at the
moment, we do not have a systematic way to describe all the possible
patterns of entanglement.  In an attempt to obtain a systematic
description of topological orders\cite{Wtoprev} in fractional
quantum Hall (FQH) states, it has been shown recently that FQH wave
functions can be classified according to their pattern of zeros,
which describes the manner in which their ground state wave
functions go to zero as the coordinates of various clusters of
electrons are brought together.\cite{WWsymm} Excited states
containing quasiparticles can also be characterized in a similar
fashion by the pattern of zeros of the wave function as electrons
are brought close to quasiparticles.  This perspective has been used
to derive some of the topological properties of quantum Hall states,
such as the types and charges of quasiparticles, in a novel
way.\cite{WWsymmqp}

However, the results in \Ref{WWsymm, WWsymmqp} are based on certain
untested assumptions, so those results need to be confirmed through other
independent methods.  Some of the FQH wave functions classified in
\Ref{WWsymm} are equal to correlation functions of certain known
conformal field theories (CFT).  For those special FQH states, we can
calculate through CFT their topological properties, such as the number of
types of quasiparticles, their respective electric charges, fusion rules, and
spins.  \cite{MR9162,WWopa} This allows us to check the results in
\Ref{WWsymm,WWsymmqp}, at least for those FQH states that are
associated with known CFTs.

In this paper, we will carry out such a calculation and compare the
results from CFT with those from the pattern of zeros.  We calculate
from CFT the total number of the quasiparticle types and their charges
for the so-called generalized and composite parafermion states (see
\eqn{noqpGP}),\cite{RR9984,WWsymm,WWsymmqp} and find agreement
with results obtained from the pattern-of-zeros approach.

We also combine the  pattern-of-zero approach and the CFT approach
to study topological properties of FQH states, and have obtained
new results that generalize those in \Ref{WWsymmqp}.
We find in general that the pattern-of-zeros approach gives rise to
a natural notion of ``translation'' that acts on quasiparticles.
This allows us to show that quasiparticles in FQH states form a
representation of a magnetic translation algebra (see \eqn{Tops}),
with members of each representation differing from each other by
Abelian quasiparticles. This is consistent with the fact that the
quasiparticles have a one-to-one correspondence with degenerate FQH
ground states on the torus, which form a representation of the
magnetic translation algebra. It also implies that various
topological properties such as fusion rules and scaling dimensions
may simplify dramatically (see \eqn{fusionRelation},
\eqn{fusionRelation1}, and \eqn{hhtlsc}). A special consequence of
this structure is that it allows us to prove quite generally that
the ground state degeneracy of FQH states on genus $g$ surfaces is
given by $\nu^{-g}$ times a factor that depends only on the
``non-Abelian'' part of the CFT and not on the filling fraction
$\nu$ (see \eqn{gsd}).

We further discuss fusion rules and their connection to domain walls
in the pattern-of-zeros sequences. There, we find a non-trivial
condition on the pattern-of-zeros of a set of quasiparticles that can
be involved in fusion with each other (see \eqn{fusionConditionSCl}).
In the general and composite
parafermion states, this condition is sufficient to completely
determine the fusion rules and may perhaps also be sufficient to do so
more generally in other FQH states.  If the latter is true, then one
can derive the fusion rules from the pattern-of-zeros.

Usually, the quasiparticle operators in a CFT are
discussed by embedding the CFT into some simpler CFT's.  The pattern-of-zeros
approach allows us to understand the structure of CFT in a more physical way.

\section{Pattern of zeros and conformal field theory}
\label{cftsec}

\subsection{FQH wave function as a correlation function in CFT}

The ground state wave function of a FQH state (in the first Landau level)
has a form
\begin{equation*}
 \Psi= \Phi(\{z_i\}) \e^{-\frac{1}{4}\sum  |z_i|^2}
\end{equation*}
where $z_i=x_i+i y_i$ is the coordinate for the $i^{th}$ electron.  Here
$\Phi(\{z_i\})$ is an antisymmetric polynomial (for fermionic electrons) or a
symmetric polynomial (for bosonic electrons).  In this paper, we will only
consider the cases of bosonic electrons where $\Phi(\{z_i\})$ is a symmetric
polynomial. The case of fermionic electrons can be included by replacing
$\Phi(\{z_i\})$  by $\Phi(\{z_i\})\prod_{i<j} (z_i-z_j)$.

In \Ref{WWsymm,WWsymmqp}, the symmetric polynomials $\Phi(\{z_i\})$
are studied and classified directly through their pattern of zeros.  In this
paper, we will study symmetric polynomials through conformal field theory
(CFT).  This is possible since for a class of ideal FQH states, the symmetric
polynomial $\Phi$ can be written as a correlation function of vertex operators
$V_\text{e}(z)$ in a CFT:\cite{MR9162,WWopa,WWHopa}
\begin{equation}
\label{PhiVe}
 \Phi(\{z_i\})=\lim_{z_\infty\to \infty} z_\infty^{2h_N}
\<V(z_\infty)\prod_i V_\text{e}(z_i) \>.
\end{equation}
Such a relation allows us to study and classify FQH states
through a study and a classification of proper CFTs.

In the above expression,
$V_\text{e}$ (which will be called an electron operator) has a form
\begin{equation*}
 V_\text{e}(z)=\psi(z)\e^{\imth \varphi(z)/\sqrt{\nu}}
\end{equation*}
where $\nu$ is the filling fraction of the FQH state.  The CFT
generated by the $V_\text{e}$ operator contains two parts.  The
first part, the simple current part, is generated by a simple
current operator $\psi$, which satisfies an Abelian fusion
rule\cite{ZF8515,GQ8723}
\begin{equation*}
 \psi_a(z)\psi_b(z) =\psi_{a+b}(z),\ \ \ \ \
\psi_a(z)\equiv [\psi(z)]^a.
\end{equation*}
The second part, the $U(1)$ ``charge'' part, is generated by the vertex
operator $\e^{\imth \varphi(z)/\sqrt{\nu}}$ of a Gaussian model, which has a
scaling dimension $h = \frac{1}{2\nu}$.  The scaling dimension  of $\psi_a$ is
denoted as $h^\text{sc}_a$.  Thus the scaling dimension of the $a^{th}$ power
of the electron operator
\begin{equation*}
V_a \equiv (V_\text{e})^a=\psi_a \e^{\imth a \varphi(z)/\sqrt{\nu}}
\end{equation*}
is given by
\begin{equation}
\label{hhsc}
 h_a=h^\text{sc}_a+\frac{a^2}{2\nu} .
\end{equation}

\subsection{The pattern-of-zeros approach and CFT approach}

In \Ref{WWsymm}, a pattern of zeros $\{S_a\}$ is introduced to
characterize a FQH state, where the integer $S_a$ is defined as
\begin{align*}
 \Phi(\{z_i\})|_{\la\to 0}=\la^{S_a}
P(\xi_1,\cdots,\xi_a;z_{a+1},\cdots)+\cdots
\end{align*}
where $z_i=\la\xi_i$, $i=1,\cdots,a$.  In other words, $S_a$ is the order of
zeros in $ \Phi(\{z_i\})$ as we bring $a$ electrons together.
The pattern-of-zero characterization also applies to FQH states
generated by CFT, so in this section we will discuss the relation between
the CFT approach and pattern-of-zero approach in a general setting.

In the pattern-of-zero approach, a FQH state is characterized by the sequence
$\{S_a\}$.  In the CFT approach, a FQH state is characterized by the sequence
$\{h_a\}$ or equivalently $\{h^\text{sc}_a\}$.  From the operator product
expansion (OPE) of the electron operators:
\begin{equation}
\label{OPEelec}
 V_a(z)V_b(w)=\frac{C_{abc}}{(z-w)^{h_a+h_b-h_c}} V_c(w)+ \cdots
\end{equation}
we find that $\{S_a\}$ and $\{h_a\}$ are closely related
\begin{equation}
\label{Sh}
 S_a=h_a-a h_1.
\end{equation}
In \Ref{WWsymm}, it was shown that $\{S_a\}$ should satisfy
\begin{align}
\label{SanyCnd}
\Del_2(a,a) &=\text{even}\geq 0,
&
\Del_3(a,b,c) &=\text{even}\geq 0  ,
\end{align}
where
\begin{align}
 \label{D2D3}
\Del_2&\equiv  S_{a+b}-S_a-S_b,
\\
\Del_3 &\equiv S_{a+b+c} -S_{a+b}-S_{b+c}-S_{a+c} +S_a+S_b+S_c  .
\nonumber
\end{align}
Finding the sequences $\{S_a\}$ that satisfy the above conditions
allows us to obtain a classification of symmetric polynomials
and FQH states.

The conditions \eq{SanyCnd} become the following conditions
on $h^\text{sc}_a$:
\begin{align}
\label{hscCnd}
 \Del_2^\text{sc}(a,b) + \frac{ab}{\nu} &= \text{integer} \geq 0,
\\
 \Del_2^\text{sc}(a,a) + \frac{a^2}{\nu} &= \text{even},
&
 \Del_3^\text{sc}(a,b,c) &=\text{even} \geq 0 ,
\nonumber
\end{align}
where
\begin{align*}
\Del_2^\text{sc} &= h^\text{sc}_{a+b}-h^\text{sc}_a-h^\text{sc}_b
\nonumber\\
\Del_3^\text{sc} &=
h^\text{sc}_{a+b+c} -h^\text{sc}_{a+b} -h^\text{sc}_{b+c} -h^\text{sc}_{a+c}
+h^\text{sc}_a +h^\text{sc}_b +h^\text{sc}_c .
\end{align*}
It is not surprising  to see that
the equations in \Eq{hscCnd} are actually a part of the defining conditions of
parafermion CFTs.\cite{ZF8515,GQ8723}
This reveals a close connection between the CFT approach and
the pattern-of-zero approach of FQH states.  This also explains why many FQH
states obtained from the pattern-of-zero construction are related to
parafermion FQH states.

After understanding the relation between the pattern-of-zero approach and the
CFT approach, we are able to consider in more detail an important issue of
stability.  In the pattern-of-zero approach, we use a sequence of integers
$\{S_a\}$ to characterize a FQH state. The question is: does the sequence
$\{S_a\}$ uniquely determine the FQH state? Can there be more than one FQH
states that give rise to the same pattern of zeros? Through a few examples, we
find that some  sequences  $\{S_a\}$ uniquely determine the corresponding FQH
states, while other  sequences  $\{S_a\}$ cannot determine the FQH state
uniquely.  Through the relation to CFT, we can address such a question from
another angle. We would like to ask: can the scaling dimensions $h_a$ of the
simple currents $V_a$ uniquely determine the correlation function of those
operators? Or more simply, can the scaling dimensions $h_a$ of the simple
currents $V_a$ uniquely determine the structure constants $C_{abc}$ in the OPE
of the simple current operators (see \eqn{OPEelec})? Such a question has been
studied partially in CFT.  It was shown\cite{ZF8515} that if
$h^\text{sc}_a=a(n-a)/n$, then $C_{abc}$ is uniquely determined. On the other
hand if $h^\text{sc}_a=2a(n-a)/n$, then $C_{abc}$ can depend on a continuous
parameter.  In this case, the pattern of zeros cannot uniquely determine the
FQH wave function. We may have many linearly independent wave functions (even
on a sphere) that have the same pattern of zeros.


\subsection{The pattern of zeros of the quasiparticle operators in CFT}

The state $\Phi_\ga$ with a quasiparticle at $\xi$ can also be expressed as a
correlation function in a CFT:
\begin{equation}
\label{PhiVeq}
 \Phi_\ga(\xi;\{z_i\})=\lim_{z_\infty\to \infty} z_\infty^{2h^q_N}
\<V_q(z_\infty)V_\ga(\xi)\prod_i V_\text{e}(z_i) \>.
\end{equation}
Here $V_\ga$ is a quasiparticle operator in the CFT which has a form
\begin{equation}
\label{Vga}
 V_\ga(z)= \si_\ga(z) \e^{\imth \varphi(z) Q_\ga/\sqrt{\nu}}
\end{equation}
where $\si_\ga(z)$ is a ``disorder'' operator in the CFT generated
by the simple current operator $\psi$.  Different quasiparticles
labeled by different $\ga$ will correspond to different ``disorder''
operators. $Q_\ga$ is the charge of the quasiparticle.

How can we obtain the properties, such as the charge $Q_\ga$, of the
quasiparticles? It is hard to proceed from the abstract symbol $\ga$
which actually contains no information about the quasiparticle. It
turns out that the pattern of zeros provides a quantitative way to
characterize the quasiparticle operator.  Such a quantitative
characterization does contain information about the quasiparticle
and will help us calculate its properties.

To obtain the  quantitative characterization,
we first fuse the quasiparticle operator with $a$
electron operators:
\begin{align}
\label{Vgaa}
 V_{\ga+a}(z) &=V_\ga V_a = \si_{\ga+a}(z) \e^{\imth \varphi(z) Q_{\ga+a}/\sqrt{\nu}}
\nonumber\\
 \si_{\ga+a}&= \si_{\ga}\psi_a,\ \ \ \ Q_{\ga+a}=Q_{\ga}+a .
\end{align}
Then, we consider the OPE of $V_{\ga+a}$ with
$V_\text{e}$
\begin{equation}
\label{VeVga}
 V_\text{e}(z) V_{\ga+a}(w)=(z-w)^{l_{\ga;a+1}} V_{\ga+a+1}(w) .
\end{equation}
Let $h_a$, $h_\ga$, and $h_{\ga+a}$ be the scaling dimensions
of $V_a$, $V_\ga$, and $V_{\ga+a}$ respectively. We have
\begin{equation}
\label{lhhh}
 l_{\ga;a+1}=h_{\ga+a+1}-h_{\ga+a} - h_1 .
\end{equation}
Since the quasiparticle wave function $\Phi_\ga(\{z_i\})$ must be a
single valued function of the $z_i$'s, $l_{\ga;a}$ must be integer.
For the wave function to be finite,  $l_{\ga;a}$ must be
non-negative. The sequence of integers $\{l_{\ga;a}\}$ gives us a
quantitative way to characterize quasiparticle operators $V_\ga$ in
CFT.  $\{l_{\ga;a}\}$ turns out to be exactly the sequence of
integers introduced in \Ref{WWsymmqp} to characterize quasiparticles
in a FQH state.  The sequence $\{l_{\ga;a}\}$ describes the pattern
of zeros for the quasiparticle $\ga$.

According to \Ref{WWsymmqp}, not all sequences $\{l_{\ga;a}\}$ describe
valid quasiparticles.  The sequences $\{l_{\ga;a}\}$ that describe valid
quasiparticles must satisfy
\begin{align}
\label{SqCond}
& S_{\ga;a+b}- S_{\ga;a} - S_b \geq 0,
\\
&S_{\ga;a+b+c}
- S_{\ga;a+b}
- S_{\ga;a+c}
- S_{b+c}
+S_{\ga;a}
+S_{b}
+S_{c} \geq 0 ,
\nonumber
\end{align}
where the integers $S_{\ga;a}$ are given by
\begin{equation*}
 S_{\ga;a}=\sum_{i=1}^a l_{\ga;i}
 =h_{\ga+a}-h_\ga - a h_1 .
\end{equation*}
The solutions of \eqn{SqCond} give us the sequences that correspond to all the
quasiparticles.

There is an equivalent way to describe the pattern of zeros $\{l_{\gamma;a}\}$
using an occupation-number sequence. Consider a one-dimensional lattice whose
sites are labeled by a non-negative integer $l$.  We can think of
$l_{\gamma;a}$ as defining the location of the $a^{th}$ electron on the
one-dimensional lattice.  Thus the sequence $\{l_{\ga;a}\}$ describes a pattern
of occupation of electrons in the one-dimensional lattice.  Such a pattern of
occupation can also be described by occupation numbers $\{n_{\gamma;l}\}$, where
$n_{\gamma;l}$ denotes the number of electrons at site $l$.  Thus, each
quasiparticle $V_{\gamma}$ defines a sequence $\{l_{\gamma;a}\}$ and an
occupation-number sequence $\{n_{\gamma;l} \}$.  The occupation-number
sequence $\{n_{\gamma;l} \}$ happens to be the same sequence that
characterizes the ground states in the thin cylinder limit for the FQH
states. \cite{SL0604,BKW0608,BH0737}

The distinct quasiparticles are actually equivalence classes of fields, where
two fields are said to belong to the same quasiparticle class (or type) if
they differ by an electron operator: $V_{\gamma} \sim V_{\gamma} V_\text{e}$.
There are a finite number of these quasiparticle classes, and this number is
an important characterization of a topological phase.  Two equivalent
quasiparticles which are related by a number of electron operators will have
nearly the same occupation-number sequence. The quasiparticle operator
$V_{\ga+b}=V_\ga V_b$ is described by
\begin{equation}
\label{lgaba}
l_{\ga+b;a}=h_{\ga+b+a}-h_{\ga+a+b-1} -h_1= l_{\ga;a+b} .
\end{equation}
Thus if two sequences $\{ l_{\gamma;a} \}$ and $\{ l_{\gamma';a} \}$ satisfy
$l_{\gamma';a} = l_{\gamma; a+b}$, then $V_{\gamma'} = V_{\gamma} V_b$ and
therefore they belong to the same quasiparticle class because they only differ
by electron operators.  Two such sequences will give occupation-number
sequences $\{ n_{\gamma;l} \}$ that are the same asymptotically as $l$ grows
large, but are different near the beginning of the sequence. Thus we can
classify the quasiparticle types by the asymptotic form of their occupation-number sequence.

Here we take the point of view that two operators are physically distinct only
if their disparity can be resolved by the electron operator. In other words,
if two operators in the conformal field theory yield the same pattern of zeros
as defined above, then the electron operator cannot distinguish between them
and therefore we identify them as the same physical operator.  This point of
view is correct if the pattern of zeros uniquely determines the correlation
functions (such as the structure constants $C_{abc}$).

Let us use $\ga=0$ to label the ``trivial'' quasiparticle created by $V_0=1$. We
see that such a trivial quasiparticle is characterized by
\begin{equation}
\label{l0hhh}
 l_{0;a+1}\equiv l_{a+1}=h_{a+1}-h_{a} -h_1.
\end{equation}
Since $h_0=0$, we see that $l_1=0$.

For the FQH states of $n$-cluster form,\cite{WWsymm,WWsymmqp}
the corresponding CFT satisfies
\begin{equation}
\label{nCnd}
\psi_n= (\psi)^n=1.
\end{equation}
As a result of this cyclic $Z_n$ structure,
the scaling dimensions of the simple currents satisfy:
\begin{equation}
 \label{hscn}
h^\text{sc}_{k n}=0,\ \ \ \ \ h^\text{sc}_{a+ n}=h^\text{sc}_{a},
\end{equation}
where $k$ is a positive integer. Let
\begin{align*}
 m &\equiv  l_{n+1} =h_{n+1}-h_1-h_{n}
.
\end{align*}
Using $
h_{n+1}-h_1-h_{n}
=\frac{(n+1)^2-n^2-1}{2\nu}=\frac{n}{\nu}
$,
we find that
the filling fraction $\nu$ is given by
\begin{equation}
\label{nunm}
 \nu=\frac{n}{m}.
\end{equation}
For such a filling fraction, we also find that
$l_{\ga;a}$ satisfies
\begin{equation}
\label{lanlam}
 l_{\ga;a+n}= l_{\ga;a}+m .
\end{equation}
This is an important consequence of the $Z_n$ structure. It implies that the
occupation numbers $n_{\gamma;l}$ are periodic: $n_{\gamma; l+m} =
n_{\gamma;l}$, with a fixed number of particles per unit cell.  From the
preceding equation it follows that the size of the unit cell is $m$ and there
are $n$ particles in each unit cell.

We also note that, according to numerical experiment\cite{WWsymm},
for $h^\text{sc}_{a}$ that satisfy \eq{hscCnd}, $m$ and $S_a$ must be even,
and the solutions satisfy
\begin{equation*}
 nh^\text{sc}_a=\text{integer} .
\end{equation*}

\subsection{Quasiparticle charge from its pattern of zeros}

Now let us calculate the quasiparticle charge $Q_\ga$ (see \eqn{Vga}) from the
sequence $\{l_{\ga;a}\}$.
Since $\si_{\ga+n}=\si_\ga$, we have (see \eq{hhsc} and \eq{lhhh})
\begin{equation}
\label{hhl}
 h_{\ga+n} - h_\ga = \frac{(Q_\ga+n)^2-Q_\ga^2}{2\nu}
=nh_1+\sum_{a=1}^n l_{\ga;a}
.
\end{equation}
Using $Q_{\ga=0}=0$, we find a formula for the charge of the quasiparticle in terms of the pattern of zeros:
\begin{equation}
\label{charge}
 Q_\ga=\frac{1}{m}\sum_{a=1}^n (l_{\ga;a}-l_a)  .
\end{equation}
which agrees with the result obtained in \Ref{WWsymmqp}.

\section{The structure of quasiparticles}

\subsection{A New Labeling Scheme}

Let $h^\text{sc}_{\ga+a}$ be the scaling dimension of $\si_\ga\psi_a$, which
satisfies
\begin{equation*}
 h^\text{sc}_{\ga+n}=h^\text{sc}_{\ga}.
\end{equation*}
Following \eq{lhhh}, we can define a new sequence  $\{l^\text{sc}_{\ga;a}\}$
that does not depend on the $U(1)$ sector of the CFT and describes the
simple-current part of the quasiparticle:
\begin{equation}
\label{lschhh}
l^\text{sc}_{\ga;a+1}=h^\text{sc}_{\ga+a+1}-h^\text{sc}_{\ga+a} - h^\text{sc}_1 .
\end{equation}
$l^\text{sc}_{\ga;a}$ has the following nice properties
\begin{equation*}
 l^\text{sc}_{\ga;a+n}=l^\text{sc}_{\ga;a},\ \ \ \ \
l^\text{sc}_{\ga+b;a}=l^\text{sc}_{\ga;a+b} .
\end{equation*}
Since $h^\text{sc}_{\ga+a} = h_{\ga+a}-\frac{(Q_\ga+a)^2}{2\nu}$,
$l^\text{sc}_{\ga;a}$ and $l_{\ga;a}$ are related:
\begin{align}
\label{lscl}
 l^\text{sc}_{\ga;a}=l_{\ga;a} -\frac{m(Q_\ga+a-1)}{n}.
\end{align}
We see that
\begin{equation*}
 nl^\text{sc}_{\ga;a}=\text{integer} .
\end{equation*}
We also see that
\begin{align*}
 h^\text{sc}_{\ga+a}&=
  h^\text{sc}_{\ga} + a h^\text{sc}_1+\sum_{b=1}^a l^\text{sc}_{\ga;b}.
\end{align*}
In particular, setting $a=n$ in the preceding equation implies that the
average over a complete period of $l^\text{sc}_{\ga;a}$ yields the scaling
dimension of the simple current operator:
\begin{equation*}
\frac{1}{n}\sum_{b=1}^n l_{\gamma;b}^{\text{sc}} = - h_1^{\text{sc}}.
\end{equation*}
It is convenient to subtract off this average to
introduce $\t l^\text{sc}_{\ga;a}$:
\begin{equation*}
 \t l^\text{sc}_{\ga;a} \equiv l^\text{sc}_{\ga;a}+h^\text{sc}_1
 =h^\text{sc}_{\ga+a+1}-h^\text{sc}_{\ga+a} ,
\end{equation*}
which also satisfies
\begin{equation}
\label{ntlscint}
 n\t l^\text{sc}_{\ga;a}=\text{integer} .
\end{equation}
We find that $\t l^\text{sc}_{\ga;a}$ satisfies
$\sum_{a=1}^n \t l^\text{sc}_{\ga;a}=0$ (see \eq{lschhh}) and
\begin{align}
\label{hhtlsc}
 h^\text{sc}_{\ga+a}&=
  h^\text{sc}_{\ga} + \sum_{b=1}^a \t l^\text{sc}_{\ga;b}.
\end{align}
We see that if $\si_{\ga'}$ and $\si_{\ga}$ are related by a simple current
operator, $\si_{\ga'}=\si_{\ga+a}=\si_{\ga} \psi_a$, then the scaling
dimension of $\si_{\ga'}$ can be calculated from that of $\si_{\ga}$ using
eqn.  (\ref{hhtlsc}).

We have seen that the different quasiparticles for an $n$-cluster FQH state
are labeled by $l_{\ga;a}$, $a=1,\cdots,n$.  In the following, we will show
that we can also use $\{Q_\ga;\t l^\text{sc}_{\ga;1},\cdots, \t
l^\text{sc}_{\ga;n}\}$ to label the quasiparticles.

Since $h_{\ga=0}^\text{sc}=0$, from \eq{hhtlsc} we see that
\begin{equation*}
 h^\text{sc}_1=\t l^\text{sc}_{0;1} \equiv \t l^\text{sc}_{1} .
\end{equation*}
Therefore, from \eq{lscl}, we see that
\begin{align*}
l_{\ga;a}
&= \t l^\text{sc}_{\ga;a}-h^\text{sc}_1 +\frac{m(Q_\ga+a-1)}{n}
\nonumber\\
&= \t l^\text{sc}_{\ga;a}-\t l^\text{sc}_1 +\frac{m(Q_\ga+a-1)}{n}.
\end{align*}
So, the quasiparticles can indeed be labeled by $\{Q_\ga;\t l^\text{sc}_{\ga;1},\cdots,
\t l^\text{sc}_{\ga;n}\}$.

We note that $\ga+1$ corresponds to a bound state between a $\ga$-quasiparticle
and an electron.  The $(\ga+1)$-quasiparticle is labeled by
\begin{equation*}
\{Q_{\ga+1};\t l_{{\ga+1};1}^\text{sc},\cdots, \t l_{{\ga+1};n}^\text{sc}\}
=
\{Q_\ga+1;\t l_{\ga;2}^\text{sc},\cdots, \t l_{\ga;n}^\text{sc},\t l_{\ga;1}^\text{sc} \}.
\end{equation*}
Since two quasiparticles that differ by an electron are regarded as
equivalent, we can use the above equivalence relation to pick an
equivalent label that satisfies $0\leq Q_\ga<1$.  For each equivalence
class, there exists only one such label.  We also see that the two
sequences $\{\t l^\text{sc}_{\ga;1},\cdots, \t l^\text{sc}_{\ga;n}\}$
for two equivalent quasiparticles only differ by a cyclic permutation.

We would like to point out that two quasiparticles with the same sequence
$\{\t l^\text{sc}_{\ga;1}, \cdots,\t l^\text{sc}_{\ga;n}\}$
but different $Q_\ga$ only differ by a $U(1)$ charge part.  This is
because $\{\t l^\text{sc}_{\ga;1}, \cdots,\t l^\text{sc}_{\ga;n}\}$ do
not depend on the $U(1)$ part of the CFT. They only depend on the
simple current part of CFT.  Using the terminology of FQH physics, the
above two quasiparticles only differ by an Abelian quasiparticle
created by inserting a few units of magnetic flux.  Inserting a unit
of magnetic flux generates a shift in the occupation number:
$n_{\ga;l}\to n_{\ga';l}=n_{\ga;l-1}$.

At this stage, and for what follows, it is helpful to see some examples as
described in Table \ref{expZ2}.  The $\nu=1/2$ $Z_2$ parafermion state has six
types of quasiparticles.  We see that the six quasiparticle types in the
$\nu=1/2$ $Z_2$ parafermions states are labeled by $\{Q_\ga;\t
l^\text{sc}_{\ga;1}, \t l^\text{sc}_{\ga;2}\}=$ $\{0;\frac12,-\frac12\}$,
$\{\frac12;\frac12,-\frac12\}$, $\{0;-\frac12,\frac12\}$,
$\{\frac12;-\frac12,\frac12\}$, $\{\frac14;0,0\}$, and $\{\frac34;0,0\}$.
$\{0;\frac12,-\frac12\}$ is the trivial quasiparticle (\ie the ground state
with no excitation).  $\{\frac12;\frac12,-\frac12\}$ is an Abelian
quasiparticle created by inserting a unit flux quantum.
$\{0;-\frac12,\frac12\}$ is a neutral fermionic quasiparticle created by
inserting two unit flux quantum and combining with an electron.
$\{\frac12;-\frac12,\frac12\}$ is the bound state of the neutral fermionic
quasiparticle with the quasiparticle created by inserting a unit flux quantum.
$\{\frac14;0,0\}$ is an non-Abelian quasiparticle.
$\{\frac34;0,0\}$ is the bound state of the above
non-Abelian quasiparticle with the quasiparticle created by inserting a unit
flux quantum.


\begin{table}[tb]
\begin{tabular}{|l|c|c|c|c||c|}
\hline
\multicolumn{6}{|c|} {$Z_2^{(1)}$  $\nu = 1$} \\
\hline
$\{Q_{\ga};l,m\}$  & $\{n_{\gamma;l}\}$
& $n \cdot \{ \tilde{l}_{\gamma;a}^{sc} \}$ & $h_{\gamma}$ & $h_{\gamma}^{sc}$ & $h_{\gamma, min}^{sc}$ \\
\hline
$\{0;0,0\}$ & 20  & 1 -1 & 0 & 0 & 0\\
\hline
$\{0;0,2\}$ & 02  & -1 1 & 1/2 & 1/2 & 0 \\
\hline
\hline
$\{1/2;1,1\}$ & 11 & 0 0 & 3/16 & 1/16 & 1/16 \\
\hline
\multicolumn{6}{ } { } \\
\hline
\multicolumn{6}{|c|} {$Z_2^{(1)}$  $\nu = 1/2$} \\
\hline
$\{Q_{\ga};l,m\}$  & $\{n_{\gamma;l}\}$ & $n \cdot \{ \tilde{l}_{\gamma;a}^{sc} \}$ & $h_{\gamma}$ & $h_{\gamma}^{sc}$ & $h_{\gamma, min}^{sc}$ \\
\hline
$\{0;0,0\}$  & 1100 & 1 -1 & 0 & 0 & 0\\
\hline
$\{1/2;0,0\}$  & 0110 & 1 -1 &  1/4 & 0 & 0 \\
\hline
$\{0;0,2\}$ & 0011 & -1 1 &  1/2 & 1/2 & 0\\
\hline
$\{1/2;0,2\}$  & 1001 & -1 1 & 3/4 & 1/2 & 0 \\
\hline
\hline
$\{1/4;1,1\}$ & 1010 & 0 0 & 1/8 & 1/16 & 1/16 \\
\hline
$\{3/4;1,1\}$ & 0101 & 0 0 & 5/8 & 1/16 & 1/16 \\
\hline
\end{tabular}
\caption{ \label{expZ2} Pattern-of-zeros sequences and scaling
dimensions defined thus far for the two simplest parafermion states:
the Pfaffian states at $\nu = 1$ and $\nu = 1/2$ with $n=2$.  The
asymptotic form of the occupation-number sequences
$\{n_{\gamma;l}\}$ in a single unit cell are listed.
$\tilde{l}_{\ga;a}^\text{sc}$ is shown for $a = 1, 2$. $\{l,m\}$ are
the $SU(2)$ labels for the parafermion primary fields; for further
explanation of the notation, see Section \ref{parafermionSect}. For
the $\nu=1$ $Z_2^{(1)}$ state, $q=1$ (where $\nu = p/q$ with $p$ and
$q$ coprime) and the three quasiparticles form a dimension 2 and
dimension 1 representation of $(\hat T_1,\hat T_2)$, as one can see
from the action of $\hat T_1$, which cyclically permutes
$\{n_{\ga;l}\}$. For the $\nu=1/2$ $Z_2^{(1)}$ state, $q=2$ and the
six quasiparticles form a dimension 4 and dimension 2 representation
of $(\hat T_1,\hat T_2)$.  (The dimension 4 representation is not an
irreducible representation. There are three irreducible
representations in both cases). }
\end{table}

\subsection{Magnetic Translation Algebra}
\label{magneticAlgebra}

We saw that the distinct quasiparticle classes can be classified by the asymptotic form of
the occupation number sequences $\{n_{\ga;l}\}$. Asymptotically, $\{n_{\ga;l}\}$
is periodic, $n_{\ga;l+m} = n_{\ga;l}$ for large $l$, so distinct
quasiparticle types can actually be classified by the asymptotic form of a single unit cell,
$\{ n_{\ga;am}, n_{\ga;am+1}, \cdots, n_{\ga;am+m-1} \} $ for large enough $a$.
Henceforth, we will drop the term $am$ in the subscript, with the understanding that
\begin{equation}
\label{nl}
| \ga \rangle \equiv \{ n_{\ga;0}, n_{\ga;1}, \cdots, n_{\ga;m-1} \}
\end{equation}
refers to the asymptotic form of a single unit cell of the occupation number sequence $\{n_{\ga;l}\}$.

In terms of the sequence \eq{nl},
there is a natural unitary operation of translation that can be defined.
In fact, we shall see that the distinct quasiparticle types, when represented using \eq{nl},
naturally form representations of the magnetic
translation algebra. We are familiar with this phenomenon in quantum Hall
systems because the Hamiltonian has the symmetry of the magnetic translation
group. Remarkably, this structure already exists in the conformal
field theory.

Let us define two ``translation'' operators $\hat{T}_1$ and $\hat{T}_2$ that
act on $\{n_{\gamma;0},  n_{\ga;1}, \cdots , n_{\ga;m-1}  \}$ in the following way:
\begin{align}
\label{Tops}
\hat{T}_1 |\ga \rangle &=
\hat{T}_1 |\{n_{\gamma;0},  n_{\ga;1}, \cdots , n_{\ga;m-1}  \} \>
\nonumber\\
&=|\{ n_{\ga;m-1}, n_{\ga;0}, \cdots, n_{\ga;m-2} \}\>  = |\ga' \rangle,
\nonumber\\
\hat{T}_2 |\ga \rangle  & =  \e^{\imth 2\pi Q_{\gamma}} |\ga \rangle.
\end{align}
Note that the label $\ga$ refers to a single representative of an entire
equivalence class of quasiparticles and that while all members of the same
class will be described by the same set of integers in \eq{nl}, their electric
charges will differ by integer units, making $e^{i 2 \pi Q_{\ga}}$ independent
of the specific representative $\ga$ and dependent only on the equivalence
class to which it belongs.

In terms of the $\{l_{\gamma;a}\}$ sequence, \eq{Tops} implies that the
sequence for $\ga'$ is closely related to that for $\ga$ plus some number $b$
of electrons:
\begin{equation}
\label{lTrans}
l_{\gamma';a} = l_{\gamma + b;a} + 1,
\end{equation}
where $b$ depends on which specific representative $\ga'$ is chosen from the equivalence class
that contains it. \eq{lTrans} implies, from \eq{charge}, that the charges $Q_{\ga} = Q_{\ga+b} - b$ and $Q_{\ga'}$
are related:
\begin{equation}
Q_{\ga'} - Q_{\ga} = \frac{1}{m} \sum_{a=1}^{n} (l_{\ga';a} - l_{\ga+b;a}) + b = \frac{n}{m} + b.
\end{equation}
This means that modulo 1, $\ga$ and $\ga'$ differ in charge by $\nu$.
From the above relations, we can
deduce that $\hat{T}_1$ and $\hat{T}_2$ satisfy the magnetic translation algebra:
\begin{equation}
\hat{T}_2 \hat{T}_1 = \hat{T}_1 \hat{T}_2 \e^{2 \pi i \nu}.
\end{equation}

The key distinction between quasiparticles in different
representations of the above magnetic algebra is that they may
differ in their non-Abelian content. They can be made of different
disorder operators $\sigma_\ga$, which are non-Abelian operators in
the sense that when $\sigma_\ga$ and $\sigma_{\ga'}$ are fused
together, the result may be a sum of several different operators. In
contrast, quasiparticles that belong to the same representation
differ from each other by only an Abelian quasiparticle.  This can
be seen as follows. For two quasiparticles $\ga$ and $\ga'$ whose
occupation-number sequences are related by a translation
$\hat{T_1}$, we have, according to \eq{lTrans}, $l_{\gamma';a} =
l_{\gamma+b;a} + 1$.  It is easily verified in this case that the
simple current part of their pattern of zeros is the same up to a
cyclic permutation: $l^{\text{sc}}_{\gamma';a} =
l^{\text{sc}}_{\gamma+b;a} = l^\text{sc}_{\ga;a+b}$, which implies
that $\ga$ and $\ga'$ are both made of the same disorder operator
$\sigma_\ga$.  It can also be verified that $\nu = (Q_{\gamma'} -
Q_{\gamma+b})$. So, modulo electron operators, the difference
between $\ga$ and $\ga'$ is solely a $U(1)$ factor.  That is, if
$\hat T_1 | \ga \rangle = | \ga' \rangle$, then the pattern of zeros
of the operator $\sigma_\ga e^{i (Q_\ga + \nu)
\sqrt{\frac{1}{\nu}}\varphi} $ is described by $|\ga' \rangle$. We
may later abuse this notation and refer to $\hat T_1$ as acting on a
quasiparticle operator $V_\ga$ to give another quasiparticle
$V_{\ga'} = V_\ga e^{i \nu \sqrt{\frac{1}{\nu}} \varphi}$, by which
we mean that $\hat T_1$ acts on the pattern-of-zeros of $V_\ga$ and
yields the pattern-of-zeros of $V_{\ga'}$.

This structure has important consequences for the topological
properties of the quasiparticles.  Let the filling fraction have a
form $\nu = p/q$ where $p$ and $q$ are coprime.  Each quasiparticle
must belong to a representation of the magnetic translation algebra
generated by $\hat T_1$ and $\hat T_2$.  The dimension of each
representation is an integer multiple of $q$ (see Table \ref{expZ2}).
This is because two quasiparticles related by the action of $\hat T_1$
differ in charge (modulo 1) by $\nu$, and therefore we come back to
the same quasiparticle if and only if we apply $\hat T_1$ a multiple
of $q$ times. The dimension of each representation is at most $m$
(where recall $ m \equiv l_{n+1}$ is the size of the unit cell of the
occupation-number sequences and $\nu = n/m$).

Let us relabel the quasiparticle $\ga$ as $(i,\alpha)$, with the
Roman index $i$ labeling the representation and the Greek index
$\alpha \in \mathbb{Z}_{c_iq}$ labeling the particular quasiparticle
within the $i^\text{th}$ representation. $c_i$ is an integer and
$c_iq$ is the dimension of the $i^{th}$ representation. $(i,
\alpha)$ and $(i, \alpha + c_iq)$ refer to the same quasiparticle.
We can choose the labels $\alpha$ such that
\begin{equation}
\hat{T}_1 |i, \alpha \rangle = |i, \alpha+1 \rangle,
\end{equation}
and this implies that the quasiparticle operator $V_{i,\alpha}$ is
related (modulo electron operators) to $V_{i,\alpha+1}$ by a $U(1)$ factor:
\begin{equation}
V_{i,\alpha+1} = e^{i \nu \frac{1}{\sqrt{\nu}} \varphi} V_{i,\alpha}.
\end{equation}
In terms of the charges, this is equivalent to writing
\begin{equation}
\label{chargeRel}
Q_{(i,\alpha+1)} \text{ mod } 1 = (Q_{(i,\alpha)} + \nu) \text{ mod } 1.
\end{equation}
Note that we consider the charge modulo one
because of the equivalence of two quasiparticles that are related by
electron operators.

In this notation, we can write the fusion rules as
\begin{equation}
V_{i,\alpha} V_{j,\beta} = \sum_{k,\ga} N_{(i,\alpha),(j,\beta)}^{(k,\ga)} V_{k,\ga}.
\end{equation}
The magnetic algebra structure of the quasiparticles implies an
important simplification in the fusion rules:
\begin{align}
\label{fusionRelation}
N_{(i,\alpha),(j,\beta)}^{(k,\ga)} = N_{(i,0),(j,0)}^{(k,\ga - \alpha - \beta)}.
\end{align}
This means that the fusion rules for all of the quasiparticles are
determined by the much smaller set of numbers given by
$N_{(i,0),(j,0)}^{(k,\delta)}$. Furthermore, since charge is conserved
in fusion, $N_{(i,0),(j,0)}^{(k,\delta)} = 0$ if $(Q_{(i,0)} +
Q_{(j,0)} - Q_{(k,\delta)}) \text{ mod } 1 \neq 0$. There are only
$c_k$ different quasiparticles in the $k^{th}$ representation that
have the same charge modulo 1, so for each $i$, $j$, and $k$, there
are actually only $c_k$ different values of $\delta$ for which
$N_{(i,0),(j,0)}^{(k,\delta)}$ must be specified. In particular,
knowing that a quasiparticle from $k$ is produced in the fusion of
$(i,0)$ and $(j,0)$ is generally not enough information to completely
specify the fusion rules. However, in some cases, even more
information can be massaged out of these relations.

The $i^{th}$ representation has dimension $c_i q$, from which it
follows that $(i, c_iq)$ and $(i,0)$ label the same quasiparticle.
From \eq{fusionRelation}, we can deduce the following identity:
\begin{equation}
\label{fusionRelation2}
N_{(i,0),(j,0)}^{(k,\delta)} =  N_{(i,c_i q),(j,0)}^{(k,\delta)} =  N_{(i,0),(j,0)}^{(k,\delta - c_iq)}.
\end{equation}
Suppose that there are integers $n$, $m$, and $l$ for which
\begin{equation}
n c_i + m c_j + l c_k = 1.
\end{equation}
This happens when the greatest common divisor (gcd) of $c_i$, $c_j$ and $c_k$ is 1. In this
case, using \eq{fusionRelation2}, one finds
\begin{equation}
N_{(i,0),(j,0)}^{(k,\delta)} = N_{(i,0),(j,0)}^{(k,\delta + q)}.
\end{equation}
This means that if one quasiparticle from the $k^{th}$ representation
is produced from fusion of $(i,0)$ and $(j,0)$, then all quasiparticles
with the same charge are also produced. In particular, if $gcd(c_i,c_j,c_k) = 1$
for all choices of $i$, $j$, and $k$, which happens when $m = q$, then the
fusion rules are completely specified by the way different representations
of the magnetic algebra fuse together. We can conclude that when $m=q$, the representations
of the magnetic algebra are all irreducible and the fusion rules decompose
in the following way:
\begin{equation}
\label{fusionDcmp}
N_{(i,\alpha),(j,\beta)}^{(k,\gamma)} = \left\{
  \begin{array}{lll}
    \bar{N}_{i,j}^k  &\mbox{ if } &( Q_{(i,\alpha)} + Q_{(j,\beta)} - Q_{(k,\gamma)}) \% 1 = 0 \\
    0 & &\mbox{ otherwise } \\
    \end{array} \right.
\end{equation}
More generally, it is straightforward to check that
\begin{equation}
\label{fusionRelation1}
N_{(i,0),(j,0)}^{(k,\delta)} = N_{(i,0),(j,0)}^{(k,\delta + gcd(c_i,c_j,c_k) q)},
\end{equation}
which implies that once $i$, $j$, and $k$ are fixed, the fusion
rules are completely specified by $gcd(c_i,c_j,c_k)$ of the fusion
coefficients. The rest of the fusion coefficients can be obtained
from \eq{fusionRelation} and \eq{fusionRelation2}. As a special,
familiar example of this, consider the Pfaffian quantum Hall states
at $\nu = 1/q$. There, the quasiparticles form two representations
of the magnetic translation algebra, one with dimension $q$, and the
other with dimension $2q$, for a total of $3q$ quasiparticles. The
quasiparticles in the dimension $2q$ representation are
$e^{il\varphi/\sqrt{q}}$ and $\psi e^{il\varphi/\sqrt{q}}$ for $l =
0,1,\cdots, q-1$. The quasiparticles in the dimension $q$
representation are of the form $\sigma
e^{i(2l+1)\varphi/2\sqrt{q}}$. $\psi$ and $\sigma$ are the primary
fields of the Ising CFT. Consider the fusion rule
\begin{equation}
\sigma e^{i\frac{2l+1}{2\sqrt{q}}\varphi} \times \sigma e^{i\frac{2l'+1}{2\sqrt{q}}\varphi} = (1 + \psi) e^{i \frac{l + l' + 1}{\sqrt{q}} \varphi}.
\end{equation}
The fact that both $1$ and $\psi$ are produced and not either one individually can
now be seen to be a special case of the analysis above: since $\text{gcd}(1,1,2) = 1$,
all quasiparticles in the dimension $2q$ representation that have the allowed charge
must be produced from the fusion of quasiparticles in the dimension $q$ representation.

\subsection{Fusion Rules, Domain Walls, and Pattern of Zeros}

The pattern-of-zeros sequences $\{l_{\gamma;a}\}$ defined thus far
are interpreted by supposing that there is a quasiparticle
$V_{\gamma}$ at the origin while electrons are successively brought
in towards it. $l_{\gamma;a}$ characterizes the order of the zero
that results in the correlation function (\ie the wave function) as
the $a^\text{th}$ electron is brought in.

Generalize this concept: imagine putting $b$ electrons at the origin
and having a sequence of integers $\{l_{b;a}\}$ that characterizes
the order of the zeros as electrons are sequentially brought in to the
origin until, after some number $a_0$ of electrons are brought in, the
quasiparticle $V_{\gamma}$ is taken to the origin and fused with the
electrons there.  We then continue to bring additional electrons in
and obtain the rest of the sequence.  In terms of the quasiparticle
sequence $\{l_{\gamma;a}\}$, the combined
sequence $\{l_{b,\ga;a}\}$ would be given by
\begin{equation}
l_{b,\gamma;a}
 = \left\{
  \begin{array}{lll}
    l_{b;a} &\mbox{ if } & a \leq a_0 \\
    l_{\gamma+b;a} & \mbox{ if } & a > a_0 \\
    \end{array} \right.
\end{equation}
If $a_0$ is large enough, the occupation-number sequence $\{ n_{l} \}$ that
corresponds to $\{l_{b,\gamma;a}\}$ will have a domain wall structure. The
first $a_0$ particles will be described by the sequence $\{ n_{b;l} \}$, while
the remaining particles will be described by the sequence $\{ n_{\gamma+b;l}
\}$.  We see that a quasiparticle not at the origin corresponds to a domain
wall between the ground state occupation distribution $\{ n_{b;l} \}$ and the
quasiparticle occupation distribution $\{ n_{\gamma+b;l} \}$.
In the large $l$ limit, $\{ n_{\gamma+b;l} \}=\{ n_{\gamma;l} \}$
and the asymptotic $\{ n_{\gamma;l} \}$ indicates that there is
a quasiparticle $\ga$ near the origin.

Extending this concept further, we see that upon bringing $V_{\gamma}$ in to
the origin after $a_0$ electrons, we can bring another quasiparticle,
$V_{\gamma'}$, in to the origin after yet another set of, say, $a_1$ electrons
have sequentially been taken to the origin. The sequence $\{l_{\ga'';a}\}$ for
$a > a_1$ will describe a new quasiparticle $\ga''$ that can be regarded as
a bound state of two quasiparticles $\ga$ and $\ga'$ near the origin.  This
suggests that by considering the sequence in such a situation, we can determine
the fusion rules of the quasiparticles.

However, the fusion of non-Abelian quasiparticles can be quite complicated, as
indicated by the fusion rule:
\begin{equation}
V_{\gamma} V_{\gamma'} \sim \sum_{\ga''} N_{\ga\ga'}^{\ga''} V_{\ga''} ,
\end{equation}
which suggests that the bound state of quasiparticles $\ga$ and $\ga'$ can
correspond to several different quasiparticles $\ga''$.
Can the consideration of the above sequence capture
such a possibility of multiple fusion channels?

The answer is yes.
Suppose $\gamma$ and $\gamma'$ can fuse to $\gamma''$. Then,
the above consideration of the fusion
of quasiparticles $\ga$ and $\ga'$ will generate
a sequence $\{ l_{b,\gamma,\gamma';a} \}$:
\begin{equation}
l_{b,\gamma,\gamma';a}
 = \left\{
  \begin{array}{lll}
    l_{b;a} &\mbox{ if } & a \leq a_0 \\
    l_{\gamma+b;a} & \mbox{ if } & a_0 < a \leq a_1 \\
    l_{\gamma''+b;a} & \mbox{ if } & a > a_1 \\
    \end{array} \right.
\end{equation}
The occupation-number sequence in this case will have two domain
walls.  For the first $a_0$ particles, the sequence will be described
by $\{ n_{b;l} \}$which corresponds to the ground state.  For the next
$a_1$ particles it will be described by $\{ n_{\gamma;l} \}$, which is
a sequence that corresponds to the quasiparticle $\ga$.  After $a_1$
it will be described by $\{ n_{\gamma'';l} \}$ which is a sequence
that corresponds to the quasiparticle $\ga''$.  In this picture,
\emph{an occupation-number sequence that contains domain walls
separating sequences that belong to different  quasiparticles
describes a particular fusion channel for several quasiparticles that
are fused together}.\cite{ABK0816}

In \Ref{ABK0816} the fusion rules for the conventional 
(Read-Rezayi) parafermion FQH states were obtained from 
the pattern of zeros by identifying the domain walls
that correspond to ``elementary'' quasiparticles. 
It is unclear whether such an approach can be applied to more general
FQH states. In the following, we will describe a very different and
generic approach that applies to all FQH states 
described by the pattern of zeros.

Notice the absence of the sequence $\{ n_{\gamma';l} \}$ in the above
consideration of the fusion $\ga\ga'\to \ga''$, even though the quasiparticle
$\gamma'$ was part of the fusion.  Here the quasiparticle $\ga'$ appears
implicitly as a domain wall between $\{ n_{\gamma;l} \}$ and $\{
n_{\gamma'';l} \}$.  This motivates us to view the fusion from a different
angle: what quasiparticle can fuse with quasiparticle $\ga$ to produce the
quasiparticle $\ga''$? From this point of view, we may try to determine $\{
n_{\gamma';l} \}$ from $\{ n_{\gamma;l} \}$ and $\{ n_{\gamma'';l} \}$ to
obtain the fusion rule.  Or more generally, the three occupation distributions
$\{ n_{\gamma;l} \}$,  $\{ n_{\gamma';l} \}$, and $\{ n_{\gamma'';l} \}$
should satisfy certain conditions if $\ga$ and $\ga'$ can fuse into $\ga''$.

\begin{figure}[tb]
\centerline{
\includegraphics[scale=0.4]{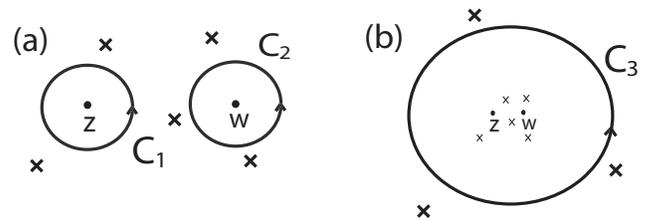}
}
\caption{
Schematic drawing of contours around which $y$ is taken. The crosses depict off-particle zeros.
(a) $y$ is taken around $C_1$ and then $C_2$, not enclosing any off-particle zeros. The
correlation function picks up a phase $2 \pi (p_{\gamma+a;b} + p_{\gamma'+c;b})$.
(b) $z$ and $w$ are brought together, fusing $V_{\gamma+a}(z)$ and $V_{\gamma'+c}(w)$, dragging
in some of the off-particle zeros along with them. $y$ is taken around the fused combination,
as illustrated by the contour $C_3$. The correlation function picks up a phase
$2 \pi p_{\gamma''+a+c;b} \geq 2\pi (p_{\gamma + a;b} + p_{\gamma'+c;b})$.
}
\label{contours}
\end{figure}

Let us now look for such a condition. Suppose two quasiparticle operators
$\gamma$ and $\gamma'$ can fuse to a third one, $\gamma''$, and consider the
OPE between the following three operators:
\begin{equation}
V_{\gamma+a}(z) V_{\gamma'+c}(w) V_b(y) \sim f(z,w,y) V_{\gamma'' + a + b + c}(z) + ...
\end{equation}
(Such an OPE makes sense if we are imagining a correlation function
with all other operators inserted at points far away from $z$, $w$, and
$y$.) Let us first fix all positions except $y$ and regard the
correlation function as a function of $y$. Zeros (poles) of the
correlation function can occur when $y$ coincides with the positions
at which other operators are inserted. However zeros can also occur at
locations away from the particles (see Figure \ref{contours}). Imagine
that we take $y$ around $w$ without enclosing any of the off-particle
zeros. The phase that the correlation function acquires upon such a
monodromy is simply $2 \pi p_{\gamma' + c;b}$. In terms of the scaling
dimensions, the integer $p_{\gamma;b}$ is given by
\begin{equation}
\label{pDef}
p_{\gamma;b}  = h_{\gamma + b} - h_{\gamma} - h_b  = p_{\gamma;b}^\text{sc} + b Q_{\gamma}/\nu ,
\end{equation}
where
\begin{equation}
p_{\gamma;b}^\text{sc} = h_{\gamma + b}^\text{sc} - h_{\gamma}^\text{sc} - h_b^\text{sc}
= \sum_{a=1}^{b} (l^\text{sc}_{\gamma;a} - l^\text{sc}_a).
\end{equation}
If we take $y$ around $z$ without enclosing any off-particle zeros, the
correlation function acquires a phase $2 \pi p_{\gamma+a;b}$. Taking $y$
around $w$ and then around $z$ thus gives a total phase of $2\pi
(p_{\gamma+a;b} + p_{\gamma'+c;b})$.  Compare that combined process with the
following process: fuse $V_{\gamma+a}$ and $V_{\gamma'+c}$ to get
$V_{\gamma''+a+c}$, and take $V_b(y)$ around $V_{\gamma''+a+c}$; physically
this corresponds to taking $z$ to be close to $w$ compared to $y$ and taking
$y$ around a contour that encloses both $z$ and $w$ (Figure \ref{contours}).
The result of such an operation is that the correlation function acquires a
phase of $2 \pi p_{\gamma''+a+c;b}$. In fusing $V_{\gamma+a}(z)$ and
$V_{\gamma'+c}(w)$ to get $V_{\gamma''+a+c}(z)$, some of the off-particle
zeros that were present before the fusion may now be located at $z$. That is,
fusing $V_{\gamma+a}(z)$ and $V_{\gamma'+c}(w)$ corresponds to taking $w
\rightarrow z$, which in the process may take some of the off-particle zeros
to $z$ as well.  Therefore we can conclude:
\begin{equation}
\label{fusionCondition}
p_{\gamma+a;b} + p_{\gamma'+c;b} \leq p_{\gamma''+a+c;b},
\end{equation}
which must be satisfied for all positive integers $a$, $b$, and $c$.  The
inequality is saturated when there are no off-particle zeros at all.

The $U(1)$ Abelian fusion rules imply charge conservation:
$Q_{\gamma} + Q_{\gamma'} = Q_{\gamma''}$, which means that the
$U(1)$ part saturates the inequality (\ref{fusionCondition}) (see
eqn. (\ref{pDef})).  This allows us to obtain a more restrictive
condition
\begin{equation}
\label{fusionConditionSC}
p_{\gamma+a;b}^\text{sc} + p_{\gamma'+c;b}^\text{sc} \leq p_{\gamma''+a+c;b}^\text{sc}
\end{equation}
which corresponds to (\ref{fusionCondition}) applied to the simple current
part. In terms of the sequences $\{ l_{\gamma;a} \}$, the condition (\ref{fusionConditionSC}) becomes
\begin{equation}
\label{fusionConditionSCl}
\sum_{j=1}^{b} (l^\text{sc}_{\gamma+a;j}  +  l^\text{sc}_{\gamma'+c;j} - l^\text{sc}_j)
\leq  \sum_{j=1}^{b} l^\text{sc}_{\gamma''+a+c;j}
\end{equation}
Remember that $l^\text{sc}_{\ga;a}$ is obtained from $l_{\ga;a}$
through \eqn{lscl} and \eqn{charge}.  The pattern-of-zero sequences
$\{ l_{\ga;a} \}$ that describe valid quasiparticles are solved from
\eqn{SqCond}.

For states that satisfy the $n$-cluster condition, the scaling dimensions and hence the
pattern of zeros have a periodicity of $n$:
\begin{equation}
p_{\gamma+a+n;b}^\text{sc} = p_{\gamma+a;b}^\text{sc} = p_{\gamma + a;b + n}^\text{sc}.
\end{equation}
Therefore, $a$, $b$, and $c$ need only run through the values $0, ..., n-1$.

The \eqn{fusionConditionSC} or \eqn{fusionConditionSCl} is the condition that
we are looking for.  The fusion coefficient $N_{\ga\ga'}^{\ga''}$ can be
non-zero only if the triplet $(\ga,\ga',\ga'')$ conserves charge,
$Q_{\ga} + Q_{\ga'} = Q_{\ga''}$, and satisfies
\eqn{fusionConditionSC} (or \eqn{fusionConditionSCl}) for any choice of
$a,b,c$.  This result allows us to calculate the fusion rules from the
pattern of zeros.

Remarkably, the condition \eqn{fusionConditionSC} or
\eqn{fusionConditionSCl} appears to be complete enough.  We find
through numerical tests that for the generalized and composite
parafermion states discussed below, condition
(\ref{fusionConditionSC}) is sufficient to obtain the fusion rules:
\it $V_{\gamma}$, $V_{\gamma'}$, and $V_{\gamma''}$ satisfy
(\ref{fusionConditionSC}) and charge conservation if and only if
$V_{\gamma}$ and $V_{\gamma'}$ can fuse to give $V_{\gamma''}$. \rm We
do not yet know, aside from these parafermion states, whether
condition (\ref{fusionConditionSC}) is sufficient to obtain the fusion
rules.  If we assume $N_{\ga\ga'}^{\ga''}=0,1$, then it is possible that
\eqn{fusionConditionSC} or \eqn{fusionConditionSCl} completely determines 
the fusion rules.

\subsection{Fusion Rules and Ground State Degeneracy on Genus $g$ Surfaces}

After obtaining the fusion rules from the pattern of zeros 
(see \eqn{fusionConditionSC}), we would like to
ask: can we check this result physically, say through numerical calculations
of some physical system? Given
a pattern-of-zeros sequence, there is a local Hamiltonian, which was
constructed in \Ref{WWsymm}, whose ground state wave function is
described by this pattern-of-zeros. The Hamiltonian can be solved
numerically to obtain quasiparticle excitations and in principle we
can check the the fusion rules. However, this approach does not really
work since the numerical calculation will produce many quasiparticle
excitations, and most of them only differ by local excitations and
should be regarded as equivalent.  We do not have a good way to determine
which quasiparticles are equivalent and which are topologically
distinct. This is why we cannot directly check the fusion rules
of the excitations through numerical calculations.

However, there is an indirect way to check the fusion rules.  The
fusion rules in a topological phase also determine the ground state
degeneracy on genus $g$ surfaces.  We can numerically compute the
ground state degeneracy on a genus $g$ surface and compare it with the
result from the fusion rules.

\begin{figure}[tb]
\centerline{
\includegraphics[scale=0.6]{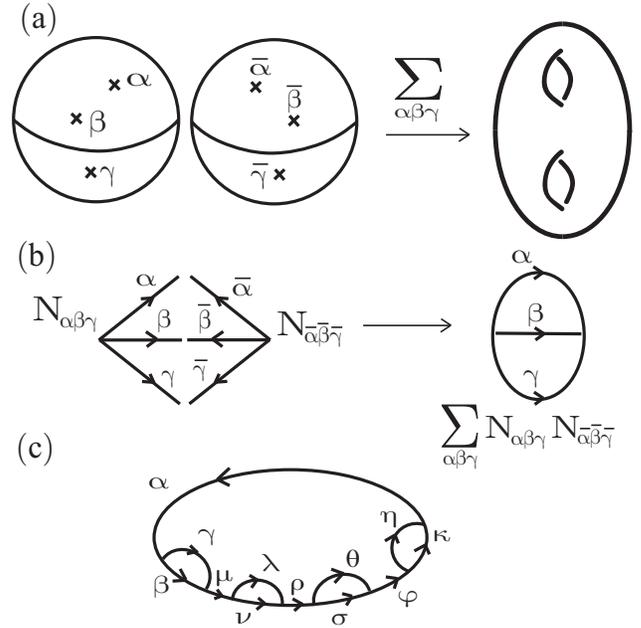}
}
\caption{ (a) Genus $g$ surfaces can be constructed by sewing together
3-punctured spheres; the case $g=2$  is shown here.  (b) 3-punctured
spheres can be depicted by vertices of $\varphi^3$ diagrams.  There is
a factor of $N_{\alpha \beta \ga}$ for each vertex with outward
directed edges $\alpha$, $\beta$, and $\ga$.  Reversing the direction
of an edge corresponds to replacing a quasiparticle with its
conjugate. The $g=2$ case is depicted by a 2-loop diagram, and has a
factor $\sum_{\alpha \beta \ga} N_{\alpha \beta \ga} N_{\bar \alpha
\bar \beta \bar\ga}$.  (c) An example of a 5-loop diagram,
corresponding to a genus $g=5$ surface.  This would give a factor
$\sum N_{\bar{\alpha} \beta \ga} N_{\bar \beta \bar\ga \mu} N_{\bar
\mu \nu \lambda} N_{\bar \lambda \bar \nu \rho} N_{\bar \rho \theta
\sigma} N_{\bar \theta \bar \sigma \varphi} N_{\bar \varphi \eta
\kappa} N_{\bar \kappa \bar \eta \alpha}$, which can be written more
compactly as $\text{Tr} \left( \sum_\alpha N_{\alpha} N_{\bar \alpha}
\right)^{g-1}$.
}
\label{gsdeg}
\end{figure}

Why do fusion rules determine the ground state degeneracy? This is
because genus $g$ surfaces may be constructed by sewing together
3-punctured spheres (see Fig. \ref{gsdeg}).  Each puncture is labeled
by a quasiparticle type, and two punctures can be sewed together by
summing over intermediate states at the punctures. This corresponds to
labeling one puncture by a quasiparticle $\ga$, labeling the other
puncture by the conjugate of $\ga$, which is referred to as
$\bar{\ga}$, and summing over $\ga$.  $\bar \ga$ is the unique
quasiparticle that satisfies $N_{\ga \bar{\ga}}^0 = 1$; the operator
that takes $\ga$ to $\bar \ga$ is the charge conjugation operator $C$:
$C_{\alpha \beta} = N_{\alpha \beta}^0$.  The dimension of the space
of states of a 3-punctured sphere labeled by $\alpha$, $\beta$, and
$\ga$ is $N_{\alpha \beta \ga} = N_{\alpha \beta}^{\bar{\ga}}$.
$N_{\alpha \beta \ga}$ is symmetric in its indices, which we can raise
and lower with the charge conjugation operator: 
\begin{align*}
N_{\alpha \beta \ga} = C_{\ga \delta} N_{\alpha \beta}^{\delta} = N_{\alpha \beta}^{\bar{\ga}} = C^{\bar \ga \delta} N_{\alpha \beta \delta}.
\end{align*}
$C^{\alpha \beta}$ is the inverse of $C_{\alpha \beta}$: $C^{\alpha \beta} C_{\beta \ga} = \delta^\alpha_\ga$.
Also, note that $C$ squares to the identity, $C_{\alpha \beta} C_{\beta \ga} = \delta_{\alpha \ga}$,
so that $C$ is its own inverse: $C_{\alpha \beta} = C^{\alpha \beta}$. 
If we represent a 3-punctured sphere by a vertex in
a $\varphi^3$ diagram with directed edges and label the outgoing
edges by $\alpha$, $\beta$, and $\ga$, each vertex comes with a
factor $N_{\alpha \beta \ga}$. A genus $g$ surface can then be
thought of as a $g$-loop diagram. This implies that the ground state
degeneracy on a torus, for example, is $\sum_{\alpha,\beta}
N_{0\alpha \beta} N_{0 \bar{\alpha} \bar{\beta}}$. The ground state
degeneracy on a genus 2 surface would be given by
\begin{equation*}
\sum_{\alpha\beta\ga} N_{\alpha \beta \ga} N_{\bar{\alpha}
\bar{\beta} \bar{\ga}} = \sum_{\alpha\beta\ga} N_{\alpha \beta}^\ga N_{\bar{\alpha} \ga}^\beta.
\end{equation*} 
In general, one obtains the following
formula for the ground state degeneracy in terms of the fusion
rules\cite{V8860} (see Figure \ref{gsdeg}):

\begin{equation}
\label{genusGDeg}
\text{ G.S.D. } = \text{Tr} \left( \sum_{a=0}^{N-1} N_{\alpha} N_{\bar{\alpha}} \right)^{g-1}.
\end{equation}
$N$ is the number of quasiparticle types, $(N_\alpha)_{\beta}^\ga = N_{\alpha \beta}^\ga$, and
matrix multiplication of the fusion matrices is defined by contracting indices, so that 
$(N_{\alpha} N_{\beta})_i^j = N_{\alpha i}^k N_{\beta k}^j$.
\eq{genusGDeg} assumes that all fields are fusing to the identity, so it applies
only when the total number of electrons is a multiple of $n$ (for $n$-cluster states).
For other cases, one must perform a more careful analysis.

We show in Appendix \ref{gsdCalculations} that \eq{genusGDeg} can be
rewritten as
\begin{equation}
\label{genusGDegQD}
\text{G.S.D.} = \left(\sum_{\ga = 0} d_\ga^2 \right)^{g-1} \sum_{\ga=0}^{N-1} d_\ga^{-2(g-1)}.
\end{equation}
$d_\ga$ is the ``quantum dimension'' of quasiparticle $\ga$. It is
given by the largest eigenvalue of the fusion matrix $N_\ga$, and it
has the property that the space of states with $n$ quasiparticles of
type $\ga$ at fixed locations goes as $ \sim d_\ga^n$ for large $n$.
In particular, Abelian quasiparticles have unit quantum dimension.  It
is remarkable that the ground state degeneracy on any surface is
determined solely by the quantum dimensions of quasiparticles.

From \eq{genusGDegQD} and the magnetic algebra structure of the
quasiparticles, we can prove that the ground state degeneracy on genus
$g$ surfaces factorizes into a part that depends only on the filling
fraction $\nu$ and a part that depends only on the simple current CFT.
In particular, we show in Appendix \ref{gsdCalculations} that
\eq{genusGDegQD} can be rewritten as 
\begin{align}
\label{gsd}
\text{ G.S.D. } &= \nu^{-g} \left( \sum_i c_i^\text{sc} d_i^{2} \right)^{g-1} \left( \sum_i c_i^\text{sc} d_i^{-2(g-1)} \right).
\end{align}
Here $\sum_i$ sums over the representations (which are labeled by
$i$) of the magnetic translation algebra. $d_i$ is the quantum dimension
of quasiparticles in the $i^{th}$ representation. $c_i^\text{sc}$ is
the number of distinct fields of the form $\psi^a \sigma_i$ for a fixed
$i$. It can be determined from the pattern-of-zeros as follows. Recall that all of the
quasiparticles in the $i^\text{th}$ representation of the magnetic
translation algebra have the same sequence
$\{l^\text{sc}_{i;a}\}$ up to a cyclic permutation.  $c_i^\text{sc}$
describes the shortest period of $\{l^\text{sc}_{i;a}\}$:
\begin{align*}
 l^\text{sc}_{i;a}=l^\text{sc}_{i;a+c_i^\text{sc}} .
\end{align*}
$\{l^\text{sc}_{i;a}\}$ always satisfies
$l^\text{sc}_{i;a}=l^\text{sc}_{i;a+n}$ and very often
$c_i^\text{sc}=n$. But sometimes, $c_i^\text{sc}$ can be a factor of
$n$. 

We see that the $c_i^\text{sc}$ are determined from the pattern of zeros
of the quasiparticles. We have seen that (under certain assumptions)
the fusion rules (and hence the quantum dimensions $d_i$) can also be
determined from the pattern of zeros.  Thus \eqn{gsd} allows us to
calculate the ground state degeneracy on any genus $g$ surface from
the pattern of zeros.

\eq{gsd} shows that the ground state degeneracy on a genus $g$
surface factorizes into $\nu^{-g}$ times a factor that depends only on
the simple current CFT. This is remarkable because $\nu^{-1}$ is
generically not an integer. The second factor may be interpreted as
the dimension of the space of conformal blocks on a genus $g$ surface
with no punctures for the simple current CFT. In particular, for genus
one, this gives $1/\nu$ times the number of distinct fields of the
form $\psi^a \sigma_i$ in the simple current CFT, a result which we
find more explicitly in the following section for the parafermion
quantum Hall states. Note that this formula assumes that the number of
electrons is a multiple of $n$; we expect a similar decomposition into
$\nu^{-g}$ times a factor that depends only on the simple current CFT
if the electron number is not a multiple of $n$, but we will not
analyze here this more complicated case.

\section{Parafermion Quantum Hall States} \label{parafermionSect}

Using the pattern-of-zero approach, we can obtain the number of types
of quasiparticles,\cite{WWsymm,WWsymmqp} the fusion rules,
\etc.  However, to obtain those results from the pattern-of-zero
approach, we have made certain assumptions. In this section, we will
study some FQH states using the CFT approach to confirm those results
obtained from the pattern-of-zero approach.

The quantum Hall states to which we now turn include the
parafermion\cite{RR9984}, ``generalized parafermion,'' and ``composite
parafermion'' \cite{WWsymm} states.  These states are all based
on the $Z_n$ parafermion conformal field theory introduced by
Zamolodchikov and Fateev\cite{ZF8515}. In the context of quantum Hall
states, we focus on the holomorphic part of the theory and leave out
the anti-holomorphic part.  The $Z_n$ parafermion CFT is generated by
$n$ simple currents $\psi_a(z)$,  $a = 0,...,n-1$, which have a $Z_n$
symmetry: $\psi_1^n = 1$ and $\psi_a = \psi_1^a$.

The field space of the theory splits into a direct sum of subspaces,
each with a certain $Z_n$ charge, labeled by $l$, with $l = 0, ..,
n-1$. The fields with minimal scaling dimension in each of these
subspaces are the so-called ``spin fields'' or ``disorder
operators'' $\sigma_l$.  Fields in each subspace are generated from
the $\sigma_l$ by acting with the simple currents: $\psi_a
\sigma_l$. Based on a relation between $SU(2)$ current algebra and
parafermion theory, a way of labeling these primary fields is as
$\Phi^l_m$.\cite{GQ8723} The spin fields are $\sigma_l = \Phi^l_l$
and the simple currents are $\psi_a = \Phi^0_{2a}$.  The $Z_n$
symmetry implies that $\Phi^l_{m+2n} = \psi_1^n \Phi^l_m =
\Phi^l_m$.  The scaling dimensions of the simple currents $\psi_a =
\Phi^0_{2a}$ are chosen to be
\begin{equation}
\label{pfScD}
\Delta^0_{2a} = \frac{a (n-a)}{n}.
\end{equation}
Such a choice then determines the scaling dimensions of the rest of the fields in the theory.
The scaling dimension $\Delta^l_m$ of the field $\Phi^l_m$ is given by
\begin{equation}
\label{parafermionWeight}
\Delta^l_m = \left\{
  \begin{array}{lll}
    \frac{l(l+2)}{4(n+2)} - \frac{m^2}{4n} + \frac{m-l}{2} &\mbox{ if } & l \leq m \leq 2n - l \\
    \frac{l(l+2)}{4(n+2)} - \frac{m^2}{4n} & \mbox{ if } &-l \leq m \leq l \\
    \end{array} \right.
\end{equation}
$l$ and $m$ satisfy
\begin{equation}
\label{lmrng}
 l+m=\text{even} ,\ \ \ \ \  0\leq l \leq n.
\end{equation}
In the $\widehat{su}(2)_n/\widehat{u}(1)$ coset formulation of the parafermion CFTs,
the following field identifications are made:
\begin{equation}
\label{fid}
\{l,m\} \sim \{n-l, m-n\} \sim \{n-l,n+m\}.
\end{equation}
Also, the $Z_n$ structure implies:
\begin{equation}
\label{fidZn}
\{l,m\} \sim \{l, m+2n\} .
\end{equation}
\begin{figure}[tb]
\centerline{
\includegraphics[scale=0.5]{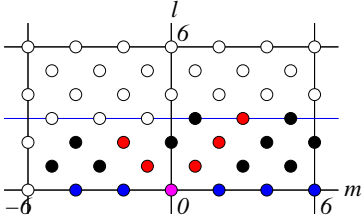}
}
\caption{
(Color online) The filled dots represent the distinct $\Phi^l_m$'s in
$Z_6$ parafermion CFT.  The blue dots represent the simple current
operators $\Phi^0_{2a}=\psi_a$ and the red dots represent the spin
operators $\Phi^l_{l}=\si_l$.
}
\label{lm}
\end{figure}
From \eq{lmrng}, \eq{fid}, and \eq{fidZn}, we can find the distinct $\Phi^l_m$'s (see Fig.
\ref{lm}).  In the pattern-of-zero picture, the identification (\ref{fid}) is
natural because quasiparticles containing the field $\Phi^l_m$ yield the same
pattern of zeros as those containing the field $\Phi^{n-l}_{n+m}$ and for this
reason are considered to be physically equivalent.

The electron operator in the conventional Read-Rezayi $Z_n$ parafermion
quantum Hall states at filling fraction $\nu$ is given by
\begin{equation}
V_\text{e}(z) = \psi_1(z) \e^{\imth \sqrt{\frac{1}{\nu}} \varphi(z)}.
\end{equation}
In the ``generalized'' $Z_n^{(k)}$ parafermion states,
we use $\psi_k$ instead of $\psi_1$ to define the electron operator:
\begin{equation}
V_\text{e}(z) = \psi_k(z) \e^{\imth \sqrt{\frac{1}{\nu}} \varphi(z)}.
\end{equation}
The states $Z_n^{(1)}$ thus correspond to the conventional Read-Rezayi states.
The condition that the electron operator have
integer or half-integer spin translates into a discrete set of possible
filling fractions for the $Z_n^{(k)}$ parafermion states:
\begin{equation}
\label{filling}
\nu = \frac{n}{nM + 2 k^2}
\end{equation}
$M$ is a nonnegative integer; that it must be nonnegative is derived from the
condition that the operator product expansion (OPE) between two electrons must
not diverge as two electrons are brought close to each other. Eqn.
(\ref{filling}) is the generalization of the well-known formula $\nu =
\frac{n}{nM + 2}$ for the conventional $Z_n$ Read-Rezayi parafermion states.
In what follows, we assume $k$ and $n$ are coprime; cases in which they are
not must be treated differently.

The quasiparticle fields take the form
\begin{equation}
V_{\gamma} = \Phi^l_m \e^{\imth Q_{\gamma} \sqrt{\frac{1}{\nu}} \varphi},
\end{equation}
where $Q_\ga$ is the electric charge of the quasiparticle. $V_{\ga}$ is a
valid quasiparticle if and only if it has a single-valued OPE with the
electron operator. To find the number of distinct quasiparticle types, we need
to find all the valid quasiparticle operators $V_{\gamma}$ while regarding two
quasiparticle operators as equivalent if they differ by an electron operator.

Since quasiparticle operators that differ by an electron operator are regarded
as equivalent, every quasiparticle is equivalent to one whose charge lies
between $0$ and $1$. Thus a simple way of dealing with this equivalence
relation is to restrict ourselves to considering operators whose charges
$Q_{\ga}$ satisfy
\begin{equation}
\label{chargeCondition}
0 \leq Q_{\gamma} < 1 .
\end{equation}
This ensures that we consider a single member of each equivalence
class, because adding an electron to a quasiparticle operator
increases its charge by one. For each primary field labeled by $\{l,
m \}$, there are only a few choices of $Q_{\ga}$ that satisfy
\eqn{chargeCondition} and that will make the operator $V_{\ga}$
local with respect to the electron operator. Finding all these
different allowed charges for each $\{ l, m\}$ will give us all the
different quasiparticle types.

The OPE between the quasiparticle operator and the electron is
\begin{align}
V_{\gamma}(z) V_\text{e}(w) \sim (z-w)^{a} \Phi^l_{m + 2k} \e^{\imth (Q_{\gamma} +1)\frac{1}{\sqrt{\nu}} \varphi},
\\
a= \Delta^l_{m+2k} - \Delta^l_m - \Delta^0_{2k} + Q_{\gamma}/\nu .
\end{align}
Locality (single-valuedness) between the quasiparticle and the
electron implies that $a$ must be an integer. Each allowed charge
$Q_{\ga}$ for a given primary field $\{ l, m \}$ can therefore be
labeled by an integer $a$ that, from \eqn{chargeCondition},
satisfies
\begin{equation}
\label{locality}
0 \leq a - \Delta^l_{m+2k} + \Delta^l_m + \Delta^0_{2k} < 1/\nu.
\end{equation}
Therefore, to find all the distinct, valid quasiparticles,
we search through all of the distinct, allowed triplets
$\{a, l, m \}$, subject to \eq{lmrng} and the identifications in \eq{fid} and \eq{fidZn},
and find those that satisfy \eqn{locality}. Carrying out this program
on a computer, we learn that the number of quasiparticles in the generalized
$Z_n^{(k)}$ parafermion states follows the formula
\begin{equation}
\label{noqp}
\mbox{ No. of Quasiparticles } = \frac{1}{2}\frac{n(n+1)}{\nu}.
\end{equation}
This is the natural generalization of the formula
$\frac{1}{2} (n M + 2) (n+1)$ that is well-known for the $k=1$ case.

The pattern-of-zeros approach has yielded not only these generalized
parafermion states, but also a series of ``composite parafermion'' states. In
these states, the relevant conformal field theory is chosen to consist of
several parafermion conformal field theories taken together, of the form
$\bigotimes_i Z_{n_i}^{(k_i)}$. We emphasize that here the $n_i$ are all
coprime with respect to one another and $k_i$ is coprime with respect to
$n_i$. Cases in which these coprime conditions do not hold should be treated
differently. Here, the electron operator is
\begin{equation}
V_\text{e} = \prod_{i=1} \psi_{k_i;n_i} \e^{\imth \sqrt{\frac{1}{\nu}} \varphi},
\end{equation}
where $\psi_{k_i;n_i}$ is a simple current of the $Z_{n_i}$ parafermion CFT.
The condition for the filling fraction, eqn. (\ref{filling}), generalizes to
\begin{equation}
\nu = \frac{N}{NM + 2N\sum_i\frac{k_i^2}{ni}},
\end{equation}
where $N = \prod_i n_i$ and $M$ is a nonnegative integer. Following a
procedure similar to that described above in the generalized parafermion case,
the condition eqn. (\ref{locality}) generalizes to

\begin{equation}
0 \leq a - \sum_i (\Delta^{l_i}_{m_i+2k_i;n_i} + \Delta^{l_i}_{m_i;n_i} + \Delta^0_{2k_i;n_i}) < 1/\nu
\end{equation}
where $\Delta^{l_i}_{m_i;n_i}$ is the scaling dimension $\Delta^{l_i}_{m_i}$ from the $Z_{n_i}$ parafermion CFT.
We find that the number of quasiparticles follows the natural generalization of eqn. (\ref{noqp}):
\begin{equation}
\label{noqpGP}
\mbox{ No. of Quasiparticles } = \frac{1}{\nu} \prod_i \frac{n_i(n_i+1)}{2}
\end{equation}
Strikingly, these results agree with the number of quasiparticles computed in an entirely different fashion through the
pattern-of-zero approach.\cite{WWsymmqp}


Since we know the fusion rules in the $Z_n$ parafermion CFTs, we can easily
examine the fusion rules in the parafermion quantum Hall states. The most
general states that we have discussed in this section have been the
$\bigotimes_i Z_{n_i}^{(k_i)}$ composite parafermion states. The
quasiparticles operators can be written as
\begin{equation}
V_\ga = \prod_{i=1} \Phi^{l_i}_{m_i;n_i} \e^{\imth \sqrt{Q_{\ga} \frac{1}{\nu}} \varphi},
\end{equation}
where $\Phi^{l_i}_{m_i;n_i}$ is the primary field $\Phi^{l_i}_{m_i}$ from the
$Z_{n_i}$ parafermion CFT.  Equivalently, we can label each quasiparticle as
$\{Q_{\ga};l_1,m_1,l_2,m_2, \cdots \}$.

The primary fields $\Phi^l_m$ in the $Z_n$ parafermion CFT enjoy the following fusion rules\cite{GQ8723}:
\begin{equation}
\Phi^l_m \times \Phi^{l'}_{m'} = \sum_{j = \mid l - l' \mid}^{min(l + l', n-l-l')} \Phi^j_{m+m'}.
\end{equation}
Therefore, in terms of the $\{ Q_{\ga}; l_1,m_1,\cdots\}$ labels, the fusion rules for the quasiparticles
in the composite parafermion states are given by
\begin{widetext}
\begin{align}
\label{FRcft}
\{Q_{\ga};l_1,m_1,\cdots\} \times  \{Q_{\ga'};l'_1,m'_1, \cdots \} =
 \sum_{l''_i = \mid l_i - l_i' \mid}^{min(l_i+l_i', n_i - l_i - l_i')}
 \{Q_{\ga}+Q_{\ga'};l''_1,m_1+m'_1,l_2'',m_2+m_2',\cdots\},
\end{align}
where there is a sum over each $l_i''$ for $i = 1, 2,\cdots$, and we make the
identifications
\begin{align}
\{Q_{\ga};\cdots, l_i, m_i, \cdots\} &\sim \{Q_{\ga};\cdots, n_i - l_i, m_i-n_i, \cdots\}
\nonumber \\
 & \sim \{Q_{\ga};\cdots, l_i, m_i+2n_i, \cdots\}
\sim \{Q_{\ga}+1;\cdots, l_i, m_i+2k_i, \cdots\}.
\end{align}
\end{widetext}
Such a fusion rule agrees with that obtained previously from the pattern of
zeros.

\section{Examples}

\begin{table}
\begin{tabular}{|l|c|c|c|c||c|}
\hline
\multicolumn{6}{|c|} {$Z_3^{(1)}$  $\nu = 3/2$} \\
\hline
$\{ Q_{\ga} ;l,m \}$  & $\{n_{\gamma;l}\}$ & $n \cdot \{ \tilde{l}_{\gamma;a}^{sc} \}$ & $h_{\gamma}$ & $h_{\gamma}^{sc}$ & $h_{\gamma, min}^{sc}$ \\
\hline
\hline
\{0; 0, 0\} & 30 &   2  0  -2   & 0 &  0 & 0\\
\hline
\{1/2; 0, -2\} & 03 &    -2  2  0   & 3/4 &  2/3  & 0 \\
\hline
\hline
\{1/2; 1, 1\}& 21&   1  -1  0   &    3/20&  1/15  & 1/15 \\
\hline
\{0; 1, 3\}&  12&    -1  0  1  &     2/15 &   2/5 &  1/15 \\
\hline
\end{tabular}
\caption{
\label{exampleZ3}
Pattern-of-zeros, scaling dimensions, and charges for the quasiparticles in
$Z_3^{(1)}$ at $\nu = 3/2$. The periodic sequence $\tilde{l}_{\gamma;a}^{sc}$
is listed for $a = 1, \cdots, 3$.  The asymptotic form of a single unit cell
of $\{n_{\gamma;l}\}$ is shown.
}
\end{table}

\begin{table}
\begin{tabular}{|l|c|c|c|c||c|}
\hline
\multicolumn{6}{|c|} {$Z_5^{(1)}$  $\nu = 5/2$} \\
\hline
\hline
$\{ Q_{\ga} ;l,m \}$  & $\{n_{\gamma;l}\}$ & $n \cdot \{ \tilde{l}_{\gamma;a}^{sc} \}$ & $h_{\gamma}$ & $h_{\gamma}^{sc}$ & $h_{\gamma, min}^{sc}$ \\
\hline
$\{ 1/2;0,6 \}$ & 05 & -2 -4 4 2 0  & $\frac{5}{4}$ & $\frac{6}{5}$ & 0 \\
\hline
$\{ 0;0,0 \} $  & 50 & 4 2 0 -2 -4 & 0 & 0 & 0\\
\hline
\hline
$\{ 1/2;1,1 \} $ & 41 & 3 1 -1 -3 0 & $\frac{3}{28}$ & $\frac{2}{35}$ & $\frac{2}{35}$ \\
\hline
$\{ 0;1,5 \}$  & 14 & -1 -3 0 3 1 & $\frac{6}{7}$ & $\frac{6}{7}$ & $\frac{2}{35}$ \\
\hline
\hline
$\{ 1/2;2,6 \}$ & 23  & -2 1 -1 2 0 & $\frac{15}{28}$ & $\frac{17}{35}$ & $\frac{3}{35}$ \\
\hline
$\{ 0;2,0 \}$ & 32 & -1 2 0 -2 1 & $\frac{2}{7}$ & $\frac{2}{7}$ & $\frac{3}{35}$ \\
\hline
\end{tabular}
\caption{
\label{exampleZ51}
Pattern-of-zeros, scaling dimensions, and charges for the
quasiparticles in $Z_5^{(1)}$ at $\nu = 5/2$. The periodic sequence
$\tilde{l}_{\gamma;a}^{sc}$ is listed for $a = 1, \cdots, 5$.
The asymptotic form of a single unit cell of $\{n_{\gamma;l}\}$ is shown.}
\end{table}

\begin{table}
\begin{tabular}{|l|c|c|c|c||c|}
\hline
\multicolumn{6}{|c|} {$Z_5^{(2)}$  $\nu = 5/8$} \\
\hline
$\{ Q_{\ga};l,m \}$  & $\{n_{\gamma;l}\}$ & $n \cdot \{\tilde{ l}_{\gamma;a}^{sc} \}$ & $h_{\gamma}$ & $h_{\gamma}^{sc}$ & $h_{\gamma, min}^{sc}$ \\
\hline
\hline
$\{ 0;0,0 \}$ & 20102000 & 6 -2 0 2 -6  & 0 & 0 & 0 \\
\hline
$\{ 5/8;0,0 \} $  & 02010200  & 6 -2 0 2 -6 & $\frac{5}{16}$ & 0 & 0\\
\hline
$\{ 2/8;0,6 \}$ & 00201020  & -6 6 -2 0 2  & $\frac{5}{4}$ & $\frac{6}{5}$ & 0 \\
\hline
$\{ 7/8;0,6 \} $  & 00020102  & -6 6 -2 0 2  & $\frac{29}{16}$ & $\frac{6}{5}$ & 0 \\
\hline
$\{ 4/8;0,2 \} $ & 20002010  & 2 -6 6 -2 0  & 1 & $\frac{4}{5}$ & 0 \\
\hline
$\{ 1/8;0,8 \} $ & 02000201  & 0 2 -6 6 -2 & $\frac{13}{16}$ & $\frac{4}{5}$ & 0  \\
\hline
$\{ 6/8;0,8 \} $  & 10200020   & 0 2 -6 6 -2 & $\frac{5}{4}$ & $\frac{4}{5}$ & 0 \\
\hline
$\{ 3/8;0,4 \} $  & 01020002 &  -2 0 2 -6 6  & $\frac{21}{16}$ & $\frac{6}{5}$ & 0 \\
\hline
\hline
$\{0; 1, 5\}$   &  01110020 &  -4  3  0  -3  4      & $\frac{6}{7}$ &  $\frac{6}{7}$ &  $\frac{2}{35}$ \\
\hline
$\{5/8; 1, 5\}$ &  00111002 &  -4  3  0  -3  4  & $\frac{131}{112}$ &  $\frac{6}{7}$ &  $\frac{2}{35}$ \\
\hline
$\{2/8; 1, 1\}$ &   20011100 &   4  -4  3  0  -3 &   $\frac{3}{28}$  & $\frac{2}{35}$ &  $\frac{2}{35}$ \\
\hline
$\{7/8; 1, 1\}$ &   02001110 &   4  -4  3  0  -3  &  $\frac{75}{112}$ & $\frac{2}{35}$ & $\frac{2}{35}$ \\
\hline
$\{4/8; 1, 7\}$ &   00200111 &   -3  4  -4  3  0  & $ \frac{6}{7}$  &  $\frac{23}{35}$ &  $\frac{2}{35}$ \\
\hline
$\{1/8; 1, 3\}$ &  10020011 &  0  -3  4  -4  3      & $\frac{75}{112}$ &  $\frac{23}{35}$ &  $\frac{2}{35}$ \\
\hline
$\{6/8; 1, 3\}$ &  11002001 &  0  -3  4  -4  3      & $\frac{31}{28}$ &  $\frac{23}{35}$ &  $\frac{2}{35}$ \\
\hline
$\{3/8; 1, -1\}$ &  11100200 &      3  0  -3  4  -4  &   $\frac{19}{112}$ &  $\frac{2}{35}$ &  $\frac{2}{35}$ \\
\hline
\hline
$\{0; 2, 0\}$   &  10101101 &  1  -2  0  2  -1  & $\frac{2}{7}$ &  $\frac{2}{7}$ &  $\frac{3}{35}$ \\
\hline
$\{5/8; 2, 0\}$ &  11010110 &  1  -2  0  2  -1  & $\frac{67}{112}$ &  $\frac{2}{7}$ &  $\frac{3}{35}$ \\
\hline
$\{2/8; 2, 6\}$ &  01101011 &  -1  1  -2  0  2      & $\frac{15}{28}$ &  $\frac{17}{35}$ &  $\frac{3}{35}$ \\
\hline
$\{7/8; 2, 6\}$ &  10110101 &  -1  1  -2  0  2      & $\frac{615}{560}$ &  $\frac{17}{35}$ &  $\frac{3}{35}$ \\
\hline
$\{4/8; 2, 2\}$ &  11011010 &  2  -1  1  -2  0      & $\frac{2}{7}$ &  $\frac{3}{35}$ &  $\frac{3}{35}$ \\
\hline
$\{1/8; 2, -2\}$    &  01101101 &  0  2  -1  1  -2      & $\frac{11}{112}$ &  $\frac{3}{35}$ &  $\frac{3}{35}$ \\
\hline
$\{6/8; 2, -2\}$    &  10110110 &  0  2  -1  1  -2  & $\frac{15}{28}$ &  $\frac{3}{35}$ &  $\frac{3}{35}$ \\
\hline
$\{3/8; 2, 4\}$ &  01011011 &  -2  0  2  -1  1      & $\frac{67}{112}$ &  $\frac{17}{35}$ &  $\frac{3}{35}$ \\
\hline
\end{tabular}
\caption{
\label{exampleZ52}
Pattern-of-zeros, scaling dimensions, and charges for the
quasiparticles in $Z_5^{(2)}$ at $\nu = 5/8$. The periodic sequence
$\tilde{l}_{\gamma;a}^{sc}$ is listed for $a = 1, \cdots, 5$.
The asymptotic form of a single unit cell of $\{n_{\gamma;l}\}$ is shown.
Note that the charges, modulo one, of two quasiparticles that
are related by a translation $\hat{T}_1$ differ
by $\nu = 5/8$, as explained in Section \ref{magneticAlgebra}.}
\end{table}

Now we will describe some specific examples of the
parafermion states, listing their pattern-of-zeros, scaling
dimensions, ground state degeneracies, and discussing their fusion rules.

In the $Z_3^{(1)}$ state at $\nu = 3/2$ (which is the bosonic $Z_3$
parafermion state\cite{RR9984}),  Table \ref{exampleZ3} shows that
there are two representations of the magnetic algebra, with two
quasiparticles in each representation. These two representations are
irreducible ($q = m$), so the fusion rules decompose as
\eq{fusionDcmp}. Labeling these two by $1$ (the identity) and
$\sigma$, we see that they satisfy the fusion rules
\begin{equation}
\sigma \sigma = 1 + \sigma.
\end{equation}
There are only two modular tensor categories of rank 2, the so-called
semion MTC and the Fibonacci MTC, and we see that these fusion rules correspond
to the Fibonacci MTC.\cite{RSW0777}

In the $Z_5^{(k)}$ states\cite{WWsymm}, we also have $q=m$ and the
quasiparticles form three irreducible representations of the
magnetic algebra.  The fusion algebra again has the simple
decomposition \eqn{fusionDcmp}.  In the $Z_5^{(1)}$ state at $\nu =
5/2$, Table \ref{exampleZ51} shows that there are two quasiparticles
in each irreducible representation, while in the $Z_5^{(2)}$ state
at $\nu = 5/8$, Table \ref{exampleZ52} shows that there are eight
quasiparticles in each irreducible representation of the magnetic
algebra. The non-trivial, non-Abelian part of the fusion rules is
given by the fusion rules among the three irreducible
representations. Labeling these three by $1$, $\sigma_1$, and
$\sigma_2$, and using \eqn{fusionConditionSC} or \eqn{FRcft}, we can
see that they satisfy the following fusion rules:
\begin{align}
\sigma_1 \sigma_1 = 1 + \sigma_2 ,
\nonumber \\
\sigma_2 \sigma_2 = 1 + \sigma_1 + \sigma_2,
\nonumber \\
\sigma_1 \sigma_2 = \sigma_1 + \sigma_2.
\end{align}
This corresponds to the $(A_1,5)_{\frac{1}{2}}$ MTC
described in \Ref{RSW0777}.

We see that when $q = m$, the decomposition of the fusion rules
\eq{fusionDcmp} into a non-trivial, non-Abelian part that depends
only on how the different irreducible representations fuse together
and a trivial Abelian part greatly simplifies these states. The
$Z_3^{(1)}$ state, which at $\nu = 3/2$ contains four
quasiparticles, has only two irreducible representations of the
magnetic algebra and therefore the non-Abelian part is described by
a simple rank 2 modular tensor category (MTC)\cite{RSW0777}: the
Fibonacci MTC. Similarly, $Z_5^{(k)}$, which for $k = 2$ and $\nu =
5/8$ has 24 quasiparticles, actually has only three different
irreducible representations of the magnetic algebra, and therefore
the non-Abelian part of its fusion rules is described by a simple
rank 3 MTC. The $Z_2$ states listed previously in Table \ref{expZ2}
have two representations, yet one of them is not irreducible. It
turns out that the non-trivial, non-Abelian part of the fusion rules
in the $Z_2$ states is described by the rank 3 Ising MTC. So, even
though $Z_5^{(2)}$ at first sight seems a great deal more
complicated than $Z_2$, their non-Abelian parts have the same degree
of complexity, namely they are both described by a rank 3 MTC.

Using the fusion rules $\eq{FRcft}$, we can also verify that the ground state
degeneracy on genus $g$ surfaces follows the decomposition $\eq{gsd}$. In particular, for the
$Z_n$ parafermion CFTs of Zamolodchikov and Fateev, the quantum dimensions
of the fields $\sigma_l \equiv \Phi^l_l$ can be found from the relation
of these theories to $SU(2)_k$ WZW models.\cite{FMS97} The result is:
\begin{equation}
d_l = \frac{\sin(\frac{\pi(l+1)}{n+2})}{\sin(\frac{\pi}{n+2})}.
\end{equation}
From the relations \eq{fid}, it follows that
for $n$ even, there are $\frac{n}{2} + 1$ distinct $\sigma_l$'s. $c_l^{\text{sc}} = n$
for $l = 0, \cdots, n/2-1$ and $c_{n/2}^\text{sc} = n/2$. For $n$ odd, there are $\frac{n+1}{2}$
distinct $\sigma_l$'s, and $c_l^\text{sc} = n$ for $l = 0, \cdots, \frac{n-1}{2}$.
Using $\eq{gsd}$, we find that the ground state degeneracy for the $Z_n^{(k)}$
states on a genus $g$ surface is given by
\begin{widetext}
\begin{equation}
\text{ G.S.D. } = \nu^{-g} n^g \times
 \left\{
  \begin{array}{lll}
    \left(\sum_{l=1}^{n/2} \sin^2(\frac{l\pi}{n+2}) + \frac{1}{2} \right)^{g-1}
\left( \sum_{l=1}^{n/2} \left(\sin(\frac{l\pi}{n+2}) \right)^{-2(g-1)} + \frac{1}{2} \right) &\mbox{ if } & $n$ \text{ is even }\\
   \left(\sum_{l=1}^{(n+1)/2} \sin^2 ( \frac{l\pi}{n+2} ) \right)^{g-1}
\left( \sum_{l=1}^{(n+1)/2} \left( \sin( \frac{l\pi}{n+2}) \right)^{-2(g-1)}\right) & \mbox{ if } & $n$ \text{ is odd }\\
    \end{array} \right.
\end{equation}
\end{widetext}
Table $\ref{GSDTable}$ lists the ground state degeneracies obtained
from the above formula for the cases $n = 2$,\cite{OKS0777} 3, 4.\cite{ABK0816} These
are the same results that one would obtain by numerically computing
\eq{genusGDeg} for the fusion rules \eq{FRcft}. For the $Z_n^{(1)}$
states, they also match the same results obtained in \Ref{ABK0816}.

\begin{table}
\begin{tabular}{|c|c|}
\hline
\multicolumn{2}{|c|} {Ground State Degeneracies for $Z_n^{(k)}$ states} \\
\hline
$n$ &  GSD \\
\hline
\hline
2 & $\nu^{-g} 2^{g-1}(2^g + 1) $\\
\hline
3 & $\nu^{-g} 3^{g} (1 + \varphi^2)^{g-1} (1+ \varphi^{-2(g-1)}) $ \\
\hline
4 & $\nu^{-g} 2^{g-1} (3^g + 1 + (2^{2g}-1)(3^{g-1} + 1))$\\
\hline
\end{tabular}
\caption{Ground State Degeneracies on genus $g$ surfaces for $Z_n^{(k)}$
parafermion quantum Hall states for the case when the number of electrons
is a multiple of $n$. $\varphi = \frac{1 + \sqrt{5}}{2}$ is the golden ratio.
\label{GSDTable}
}
\end{table}

\section{Summary}

Motivated by the characterization of symmetric polynomials and FQH
states through the pattern of zeros,\cite{WWsymm} we examined the
CFT generated by simple currents in terms of the pattern of zeros.
This reveals a deep connection between the simple current CFT and
FQH states.  The point of view from the pattern of zeros reveals a
magnetic translation algebra that acts on the quasiparticles.  This
allows us to greatly simplify the fusion algebra of the quasiparticles.  
It also allows us to show
in general that the ground state degeneracy on genus $g$ surfaces is
$\nu^{-g}$ times a factor that depends only on the simple current
part of the CFT. More importantly, we are able to derive a necessary
condition on the fusion rules of the quasiparticles based on the
pattern of zeros.  Such a necessary condition is sufficient to
produce a full set of fusion rules for quasiparticles if we assume
$N^{\ga''}_{\ga\ga'}=0,1$.

The results obtained from the pattern-of-zeros approach is checked against
known results from the generalized and composite parafermion CFT.  In
particular, we find that the number of quasiparticles obtained from the
pattern-of-zeros approach\cite{WWsymmqp} agrees with that obtained from the
CFT approach.  The fusion rules obtained from the pattern of zeros also agrees with the
result from the CFT calculation for generalized and composite parafermion FQH
states.  Those agreements further demonstrate that the pattern-of-zeros
approach is quite a powerful tool to characterize and to calculate the
topological properties of generic FQH states.

We would like to thank Zhenghan Wang for many helpful discussions.  This
research is supported by NSF Grant DMR-0706078.

\appendix

\section{Scaling Dimensions of Quasiparticles}


The way the magnetic algebra structure of the quasiparticles factorizes in the fusion rules
is also seen in another topological property of the quasiparticles:
the scaling dimensions, or spins, of the quasiparticles. Since quasiparticles that belong to the same
representation of the magnetic algebra are described by sequences
$\{l^\text{sc}_{\gamma;a}\}$ that are related by a cyclic permutation,
from \eqn{hhtlsc} we see
that for each irreducible representation there is a single number
$h_{\gamma;\text{min}}^\text{sc}$ that we need in order to calculate the
scaling dimensions of all other quasiparticles in the same representation.
$h_{\gamma;\text{min}}^\text{sc}$ is the minimum of
$h_{\gamma+a}^\text{sc}$ over all the quasiparticles that belong to this
representation.  Given $h_{\gamma;\text{min}}^\text{sc}$, the scaling
dimension of the quasiparticle $V_{\gamma + a}$ can be calculated from its
pattern of zeros $\{ \t l^\text{sc}_{\gamma+a;b} \}$.

It is not obvious that the information to obtain $h_{\gamma;\text{min}}^\text{sc}$  for
each irreducible representation is even contained in the pattern of zeros. It may
be that $\{ l_{\gamma;a} \}$ is not enough information to uniquely specify the CFT
and therefore also not enough information to completely determine the scaling dimensions of
the fields that are contained in the theory.
However, in the case where the pattern of zeros corresponds to
the (generalized and composite) parafermion states discussed above, there are
explicit formulas in terms of $l_{\ga;a}$ that yield  $h_{\gamma;\text{min}}^\text{sc}$.
This comes as no surprise because in these cases, the pattern-of-zeros completely
specifies the CFT. We do not have a formula that can even be
applied in the more general situations.

Let us now describe how to calculate the scaling dimension $h_{\gamma+a}$
of the operator $V_{\gamma+a}$ given $h_{\gamma;\text{min}}^\text{sc}$.
First we must find the index $a_0$ at which
\begin{equation}
h_{\gamma;\text{min}}^\text{sc} = h^\text{sc}_{\gamma+a+a_0}.
\end{equation}
This is equivalent to finding the index $a_0$ at which
\begin{equation}
h_{\gamma+k} - h_{\gamma+a+a_0} \geq 0
\end{equation}
for all $k$, because $h_{\gamma;\text{min}}^\text{sc}$ is defined to be the minimum
of $h_{\gamma+k}$ over all $k$. Recalling that
$\t l^\text{sc}_{\gamma+a;b+1} = h^\text{sc}_{\ga+a+b+1} - h^\text{sc}_{\ga+a+b}$,
we see that $a_0$ is therefore the index at which
\begin{align}
\label{a0cond}
\sum_{i=1}^{k} \t l^\text{sc}_{\ga+a;a_0 + i}
&= h^\text{sc}_{\ga + a + a_0 + k} - h^\text{sc}_{\ga + a + a_0} &
\nonumber \\
&= h^\text{sc}_{\ga + a + a_0 + k} - h^\text{sc}_{\ga;\text{min}} &\geq 0
\end{align}
for all $k$. Using \eqn{a0cond}, we can determine $a_0$ from
$\{\t l^\text{sc}_{\ga;a} \}$, after which we can determine $h_{\gamma+a}$ using
\eqn{hhtlsc}:
\begin{equation}
h_{\ga+a}^\text{sc} = h_{\gamma;\text{min}}^\text{sc} - \sum_{b=1}^{a_0} \t l^\text{sc}_{\ga + a; b}.
\end{equation}

\section{Ground State Degeneracy on Genus $g$ Surfaces}
\label{gsdCalculations}

Here we illustrate how \eq{genusGDegQD} and \eq{gsd} can be determined from \eq{genusGDeg}.
First we observe that the fusion matrices $N_{\alpha}$
commute (and can therefore be simultaneously diagonalized) because the fusion of any three quasiparticles
$\alpha$, $\beta$, and $\ga$ should be independent of the order in which 
they are fused together. Remarkably, there is a symmetric unitary matrix $S$, known as the modular $S$ matrix,
which squares to the charge conjugation operator, $S_{\alpha \beta} S_{\beta \ga} = C_{\alpha \ga}$,
and that simultaneously diagonalizes all of the fusion matrices:\cite{V8860, MS8851, MS8977}
\begin{equation}
\label{fusionDiag}
N_{\alpha\beta}^\ga = \sum_n S_{\beta n} \lambda_{\alpha}^{(n)} S^\dagger_{\gamma n}.
\end{equation}
Using \eq{fusionDiag} and the fact that $N_{\alpha 0}^\beta =  \delta^\beta_\alpha$, the
eigenvalues $\lambda_{\alpha}^{(n)}$ of the fusion matrix $N_{\alpha}$
can be written in terms of $S$:
\begin{equation}
\lambda_\alpha^{(n)} = \frac{S_{\alpha n}}{S_{0n}}.
\end{equation}
$S$ also has the remarkable property that the largest eigenvalue of
$N_\alpha$, which is the quantum dimension $d_\alpha$, is given by
$\lambda_\alpha^{(0)}$:
\begin{equation}
\label{qd}
d_\alpha = \frac{S_{0\alpha}}{S_{00}}.
\end{equation}
Inserting \eqn{fusionDiag} into \eqn{genusGDeg} yields
\begin{align}
\text{ G.S.D. } &= \sum_{n=0}^{N-1} \left( \sum_{\alpha=0}^{N-1} \lambda_{\alpha}^{(n)} \lambda_{\bar \alpha}^{(n)} \right)^{g-1}
\nonumber \\
& = \sum_{n=0}^{N-1} (S_{0n})^{-2(g-1)} \left( \sum_{\alpha=0}^{N-1} S_{\alpha n} S_{\bar \alpha n} \right)^{g-1}.
\end{align}
Using the fact that $S_{\alpha \beta}S_{\beta \ga} = C_{\alpha \ga}$ and 
$C_{\alpha \beta} C_{\beta \ga} = \delta_{\alpha \ga}$, we see that 
$\sum_{\alpha} S_{\alpha n } S_{\bar \alpha n} =  \sum_{\alpha \beta} S_{n \alpha} C_{\alpha \beta} S_{\beta n}  = 1$, 
so that  \eq{genusGDeg} can be rewritten as
\begin{align}
\label{genusGDegQDApp}
\text{ G.S.D. } = \sum_{\ga=0}^{N-1} S_{0\ga}^{-2(g-1)}
 = \left(\sum_{\ga = 0} d_\ga^2 \right)^{g-1} \sum_{\ga=0}^{N-1} d_\ga^{-2(g-1)},
\end{align}
where in the last equality we use the fact that
$\sum_{\alpha} d_{\alpha}^2 = S_{00}^{-2}$,
which follows from \eq{qd} and $\sum_{\alpha} S_{0\alpha}S_{\alpha 0} = 1$.

From the magnetic algebra structure of the quasiparticles that was
described in Section \ref{magneticAlgebra}, we know that quasiparticles
in the same representation of the magnetic algebra differ from each 
other by Abelian quasiparticles and thus they all have the same quantum
dimension. Since there are $c_i q$ quasiparticles in the $i^{th}$ representation,
we can see that \eq{genusGDegQDApp} can be rewritten as
\begin{align}
\label{genusGDegQDcd}
\text{ G.S.D. } = q^g \left( \sum_i c_i d_i^{2} \right)^{g-1} \left( \sum_i c_i d_i^{-2(g-1)} \right),
\end{align}
where the sum over $i$ is a sum over the different representations of the magnetic algebra
and, as defined in Section \ref{magneticAlgebra}, $c_iq$ is the dimension of the $i^{th}$
representation. Recall that $\nu = p/q$ with $p$ and $q$ coprime. $d_i$
is the quantum dimension of the quasiparticles in the $i^{th}$ representation.

To proceed further, let us pause to consider the structure of the simple current CFT. The simple current CFT
contains the ``disorder'' fields $\sigma_i$, which are primary with respect to the
algebra generated by the simple current $\psi(z)$. There are also fields of the form
$\psi^a \sigma_i$, which are primary with respect to the Virasoro algebra.
Since $\psi^n = 1$, there are at most $n$ different fields of the form $\psi^a \sigma_i$.
However, these fields are not necessarily all distinct. It may be the case that $\sigma_i$
and $\psi^a \sigma_i$ refer to the same field for certain values of $a$. This occurs when
these two fields have the same pattern-of-zeros sequences. That is, when
\begin{equation}
l_{i;b}^\text{sc} = l_{i;a+b}^\text{sc}
\end{equation}
for $b = 0, \cdots, n-1$. Let us suppose that this happens when $a$ is a multiple
of some integer $c_i^{sc}$. Then, $\psi^{c_i^\text{sc}} \sigma_i$ and $\sigma_i$ label
the same fields and so there are only $c_i^{\text{sc}}$ distinct fields
of the form $\psi^a \sigma_i$. Note that $c_i^\text{sc}$ must divide $n$.

Now recall that the action of $\hat T_1$ on some quasiparticle
$V_{i,\alpha} = \sigma_i e^{i Q_{(i,\alpha)} \sqrt{\frac{1}{\nu}} \varphi}$
is to take it to a new quasiparticle that differs from $V_{i,\alpha}$ by a $U(1)$ factor:
\begin{equation}
\hat T_1: \sigma_i e^{i Q_{(i,\alpha)} \sqrt{\frac{1}{\nu}} \varphi} \rightarrow \sigma_i e^{i (Q_{(i,\alpha)} + \nu) \sqrt{\frac{1}{\nu}} \varphi}.
\end{equation}
So if we apply $\hat T_1^{c_i q}$ to $V_{i,\alpha}$, we get
\begin{align}
\hat T_1^{c_i q}: V_{i,\alpha} &\rightarrow \sigma_i e^{i (Q_{(i,\alpha)} + c_i q\nu) \sqrt{\frac{1}{\nu}} \varphi}
\nonumber \\
&= \sigma_i e^{i (Q_{(i,\alpha)} + c_i p) \sqrt{\frac{1}{\nu}} \varphi}
\nonumber \\
& \sim \psi^{n - c_i p} \sigma_i e^{i Q_{(i,\alpha)} \sqrt{\frac{1}{\nu}} \varphi},
\end{align}
where in the last step we have used the fact that two quasiparticles are equivalent
if they differ by electron operators. Since $c_iq$ is the dimension of
the $i^{th}$ representation, quasiparticles in the $i^{th}$
representation are invariant under the action of $\hat T_1^{c_i q}$. This means that
$
\sigma_i e^{i Q_{(i,\alpha)} \sqrt{\frac{1}{\nu}} \varphi} \sim \psi^{n - c_i p} \sigma_i e^{i Q_{(i,\alpha)} \sqrt{\frac{1}{\nu}} \varphi}
$.
This happens only when $\sigma_i$ and $\psi^{n - c_i p} \sigma_i$ refer to the same
fields. Since $\sigma_i \sim \psi^{c_i^\text{sc}} \sigma_i \sim \psi^{n-c_i^\text{sc}} \sigma_i$
all refer to the same field, we find that
\begin{equation}
c_i p = c_i^\text{sc}.
\end{equation}
Inserting this in \eq{genusGDegQDcd} yields \eq{gsd}.


\end{document}